\documentclass[fleqn,usenatbib]{mnras}

\usepackage{newtxtext,newtxmath}

\usepackage[T1]{fontenc}
\usepackage{ae,aecompl}

\usepackage{graphicx}	
\usepackage{amsmath}	
\usepackage{amssymb}	

\usepackage{hyperref}
\usepackage{algorithm}
\usepackage[noend]{algpseudocode}

\usepackage{soul}

\DeclareMathOperator*{\argmax}{argmax}
\newcommand*\Let[2]{\State #1 $\gets$ #2}

\title[Real-time colouring and filtering with graphics shaders]{Real-time colouring and filtering with graphics shaders}

\author[D. Vohl et al.]{
D. Vohl,$^{1,2}$\thanks{E-mail: dvohl@swin.edu.au}
C. J. Fluke,$^{1,2}$
D. G. Barnes,$^{3,4}$
A. H. Hassan$^{1}$\\
$^{1}$Centre for Astrophysics \& Supercomputing, Swinburne University of Technology, 1 Alfred Street, Hawthorn 3122, Australia\\
$^{2}$Advanced Visualisation Laboratory, Digital Research \& Innovation Capability Platform, Swinburne University of Technology, Hawthorn, Australia\\
$^{3}$Monash e-Research Centre, Monash University, 14 Alliance Lane, Clayton 3168, Australia\\
$^{4}$Faculty of Information Technology, Monash University, Clayton, Victoria, Australia
}

\date{Accepted XXX. Received YYY; in original form ZZZ}

\pubyear{2017}

\begin{document}
\label{firstpage}
\pagerange{\pageref{firstpage}--\pageref{lastpage}}
\maketitle

\begin{abstract}
Despite the popularity of the Graphics Processing Unit (GPU) for general purpose computing, one should not forget about the practicality of the GPU for fast scientific visualisation. As astronomers have increasing access to three dimensional (3D) data from instruments and facilities like integral field units and radio interferometers, visualisation techniques such as volume rendering offer means to quickly explore spectral cubes as a whole. As most 3D visualisation techniques have been developed in fields of research like medical imaging and fluid dynamics, many transfer functions are not optimal for astronomical data. We demonstrate how transfer functions and graphics shaders can be exploited to provide new astronomy-specific explorative colouring methods. We present 12 shaders, including four novel transfer functions specifically designed to produce intuitive and informative 3D visualisations of spectral cube data. We compare their utility to classic colour mapping. The remaining shaders highlight how common computation like filtering, smoothing and line ratio algorithms can be integrated as part of the graphics pipeline. We discuss how this can be achieved by utilising the parallelism of modern GPUs along with a shading language, letting astronomers apply these new techniques at interactive frame rates. All shaders investigated in this work are included in the open source software {\tt shwirl} \citep{Vohl2017ascl.soft04003V}.
\end{abstract}




\section{Introduction}\label{sec::introduction}

A spectral cube is a multidimensional array, from which the principal three dimensions are two spatial dimensions and a spectral or a velocity dimension. This three-dimensional (3D) representation permits the investigation of various features of a source (e.g. galaxies, planetary nebulae), like complex velocity and kinematic structures, or spatially resolved emission lines. 

Astronomers have an increasing access to spectral cubes from: radio telescopes and radio interferometers [e.g. Atacama Large Millimeter Array \citep[ALMA; e.g.][]{Whitmore2014ApJ...795..156W},  Australian SKA Pathfinder \citep[ASKAP; ][]{Johnston2008ExA} and the APERTIF upgrade to the Westerbork Synthesis Radio Telescope\citep[][]{Verheijen2009pra..confE..10V}]; integral field units [IFU; e.g. KMOS \citep{Sharples2004SPIE.5492.1179S}, the MaNGA Integral Field Unit on the Sloan Telescope  \citep{DroryEtAl2015AJ}, and the Multi Unit Spectroscopic Explorer \citep[MUSE, ][]{Bacon2010SPIE.7735E..08B}
]; and imaging Fourier transform spectrometers [e.g. SITELLE \citep{Martin2016MNRAS.463.4223M}]. 

Upcoming large-scale surveys, like the many planned neutral hydrogen (HI) surveys \citep[e.g.][]{Verheijen2009pra..confE..10V, KoribalskiHarveySmith2009WALLABY}, will generate terabytes of new spectral cube data on a daily basis. Within these large cubes, many hundreds of galaxies are expected to be detected, producing sub-cubes for further analysis and investigation.

The complex 3D nature of these sources \citep{Sancisi2008A&ARv..15..189S}, and the low signal-to-noise characteristics of the data, makes it difficult to develop a fully automated and reliable pipeline \citep{Popping2012PASA...29..318P, Punzo201745}. Therefore, there is a need to explore new visualisation techniques that reduce the exploration period by visually enhancing physically meaningful features in the data. Moreover, these techniques should be part of fast computational solutions that allow processing of a large amount of data in a reasonable amount of time. 

\subsection{Two-dimensional techniques}
Historically, spectral cubes have primarily been visualised with two dimensional (2D) techniques. This can be attributed to their ease of use for scientific publication in paper form, and the simplicity of computation. 

One classic 2D visualisation technique involves the generation of a movie-like sequence, where each slice of the spectral or velocity dimension of a cube is rendered one after the other. As pointed out early on by  \citet{Norris1994ASPC...61...51N} and \citet{Oosterloo1995PASA...12..215O}, conventional movie techniques require too much time for the eyes and brain to associate information in different velocity components of the spectral cube --- limiting the ability to gain an intuitive impression of the data as a whole. The static counterpart of the movie method, known as channel map visualisation, consists of plotting the same slices individually to evaluate structures at a given wavelength or velocity bin \citep[e.g.][]{Borkin2005astro.ph..6604B}. 

Alternatives exist that combine channel maps into a single view. One option is the {\em renzogram}\footnote{See for example the renzogram routine of the Groningen Image Processing System (GIPSY) : \url{https://www.astro.rug.nl/~gipsy/tsk/renzogram.dc1} (last accessed 27 March 2017).}, named after its pioneer, Renzo Sancisi. A renzogram displays kinematic and spatial information in a single contour plot. Using one colour-coded contour per channel, multiple peaks or sub-structures can be seen in the velocity profile.

The more widely used alternative combines channels, computing derived 2D maps based on statistical moments \citep[e.g. ][]{Walter2008AJ}. In a moment map, a pixel at coordinate (x,y) represents a statistical quantity obtained from the spectrum elements at spatial coordinate (x,y) in the spectral cube. Moment maps provide a condensed view of physical quantities like the overall intensity or flux distribution (0th moment), velocity field (1st moment) and velocity dispersion (2nd moment). 

A drawback of using static, pre-computed moment maps for data exploration is the lack of interactivity with the original data.
Instead, it is desirable to expose the expert to a complete dataset, while providing a realtime response. The full 3D view of a source simultaneously shows both its flux distribution and its spectral or kinematic properties, displaying an immediate overview of the structures and coherence in the data \citep{Oosterloo1995PASA...12..215O, Goodman2012, Hassan2013MNRAS.429.2442H, PunzoEtAl2015A&C}. 

\subsection{Volume rendering}

Volume rendering permits a spectral cube to be visualised interactively, as a whole, and from arbitrary view-points \citep[e.g.][]{gooch1995astronomers, Barnes2006PASA...23...82B, Hassan2013MNRAS.429.2442H, PunzoEtAl2015A&C, Taylor2015A&C....13...67T, Ferrand2016arXiv160708874F, Punzo2016A&C....17..163P, Vohl10.7717/peerj-cs.88}. 

Volume rendering projects a 3D scalar field (e.g. 3D array) onto a 2D plane, or {\em image plane}, producing an image to be rendered on a display device. From its ability to provide a global 3D view of the data, volume rendering plays a role in the discovery of new phenomena, unexpected relations, or previously unidentified patterns that are deemed difficult to be accomplished with automated techniques \citep{Beeson2004SPIE.5493..242B, Goodman2012, Punzo2016A&C....17..163P}.

Coupled with the computational power of modern Graphics Processing Units (GPUs), volume rendering methods offer the possibility to explore and manipulate data in realtime -- a step forward for the development of next-generation visualisation and analysis software. This approach opens new ways to: dynamically explore spectral cubes; 
rapidly compute 2D moment maps and their 3D equivalents -- presented here for the first time; or  compute voxel-by-voxel operations on multiple spectral cubes, such as spectral line ratios from multi-wavelength observations. 

\subsection{New directions: transfer functions \& graphics shaders}
Data exploration and discovery often requires pre-processing of data (e.g. filtering and smoothing) as a separate stage of a multi-part analysis and visualisation workflow. With this processing model, it can be difficult to immediately assess the impact of a particular set of parameters values --- often requiring a time consuming sequence of trial and error. This represents a more significant bottleneck for cases where multiple files have to be explored. Instead, it is desirable to couple interactive parameter selection with real-time visual feedback. 


The hardware architecture of modern commodity GPUs allows greater coordination of analysis and visualisation through a shared memory space.  In practice, we can provide new visual representations of spectral cube data, linked to voxel-based analysis tasks, through the use of {\em transfer functions} and {\em graphics shaders}.

A transfer function is an arbitrary function that combines voxels' properties to set the colour, intensity, or transparency level of each pixel in the final image. Transfer functions have an important impact in the process of scalar data visualisation, as the use of colour helps the human brain to gain an understanding of the data by emphasising some features while suppressing others. 

A graphics shader (hereafter shader) is an algorithmic kernel used to compute several properties of the final image such as colour, depth, and/or transparency. Shaders are particularly suited to computing transfer functions, and are an integral part of the graphics pipeline on GPUs.  

While this is an active field of research in other application fields like medical imaging \citep[e.g.][]{ArensVG:VG10:077-083, LjungLKGHHY16}, there has not yet been any systematic investigation of the use of transfer functions and shaders in astronomy [see \citet{gooch1995astronomers} for early work]. \citet{HassanFluke2011PASA} recommended that the development of customized transfer functions and shaders should be a priority for next generation visualisation tools in astronomy. 


\subsection{Overview}

This article investigates how transfer functions and shaders can be exploited to provide new, astronomy-specific, explorative colouring methods. 

We present 12 shaders, including four novel transfer functions specifically designed to produce intuitive and informative 3D visualisations of spectral cube data. We compare their utility to classic colour mapping. 

The remaining shaders highlight how common filtering and smoothing algorithms --- like dynamic histograms and box or Gaussian smoothing --- or computing an emission line ratio can be integrated as part of the graphics pipeline to meaningfully modify 3D data at interactive rate.

We discuss how utilising an interactive shading language, along with the parallelism of modern GPUs, provides speed and control over the visual outcome. While the proposed techniques focus on volume rendering of spectral cube data, their application scope can be extended to other 3D volume data, like N-body and hydrodynamic simulations. Our investigation suggests that custom transfer functions and shaders can have an important role in the development of future visualisation and analysis astronomical software. 

The remainder of the paper is structured as follows. In Section \ref{sec::background-concepts}, we introduce background concepts about common visualisation techniques of spectral cube data. In Section \ref{sec::transfer_functions}, we introduce our advanced colouring techniques by describing new transfer functions in the form of shaders. In Section \ref{sec::Filters}, we explain how common computation of filtering, smoothing, and line ratio algorithms can be introduced to the graphics pipeline in order to be used in real-time. 
In Section \ref{sec::results_discussion}, we demonstrate, compare, and discuss the visual outcome of all transfer functions. In Section \ref{sec::performance}, we report on the performance of each smoothing kernel using a range of GPUs, including remote deployments using cloud computing. Finally, we discuss future work and conclude in Section \ref{sec::conclusion}.

\section{Background}
\label{sec::background-concepts}
\subsection{Volume rendering and applications in astronomy}
\label{sec::volume-rendering}
\begin{figure}
\includegraphics[width=8.3cm]{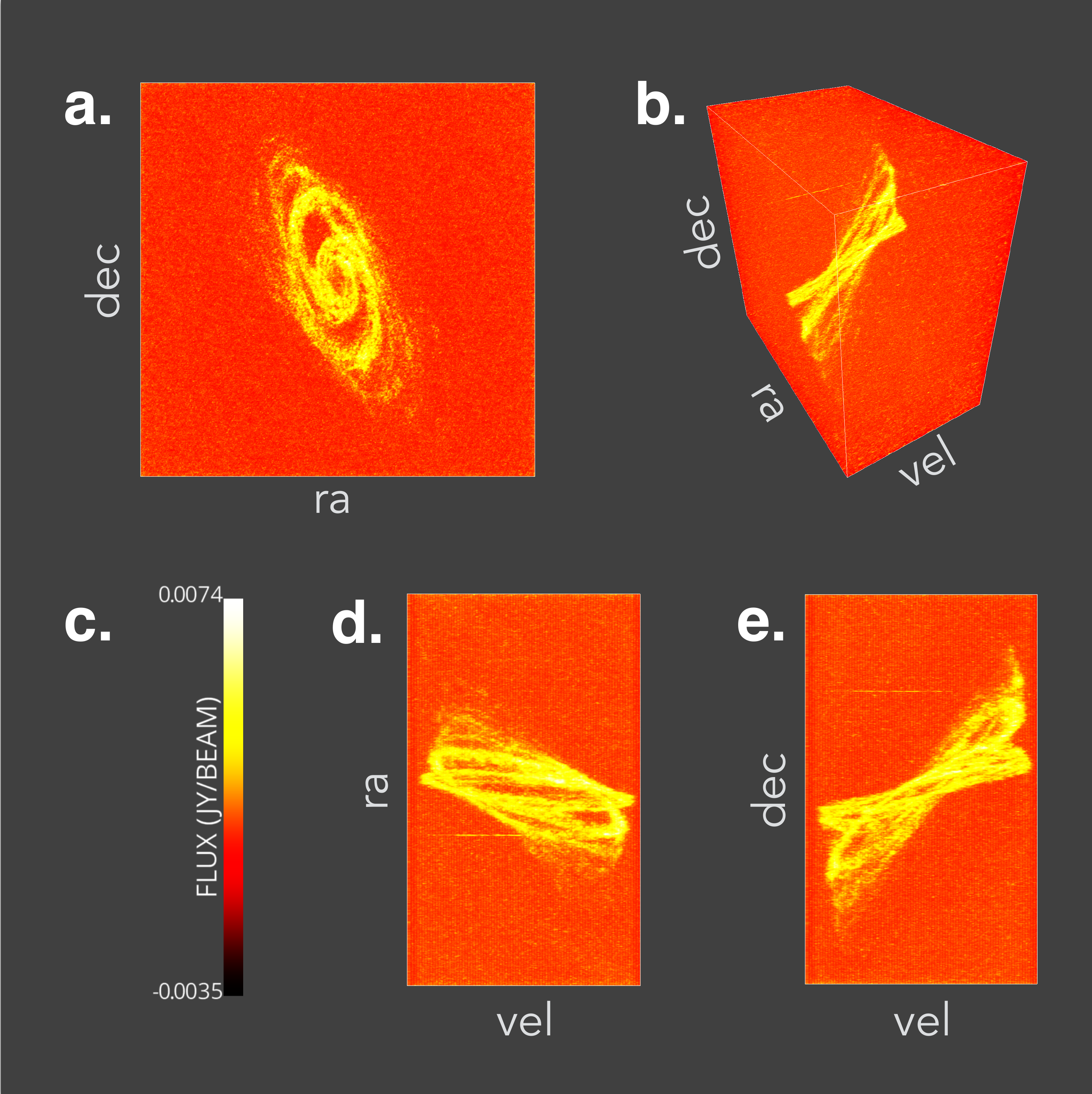}
\caption{Different points-of-view of a spectral cube --- showing galaxy NGC 2903 from The HI Nearby Galaxy Survey (THINGS) --- rendered using ray-tracing volume rendering: {\bf a.} spatial view showing right ascension (ra) and declination (dec); {\bf b.} a blend of spatial and spectral views showing ra, dec and velocity (vel); {\bf c.} a colour map used to map spectral flux density [in unit of jansky per beam (Jy/beam)] to colour; {\bf d.} spectral view (vel as a function of ra); {\bf e.} spectral view (vel as a function of dec). Panels a. d. and e. are using a parallel projection, while panel b. is using a perspective projection. All volume renderings use the Maximum Intensity Projection transfer function.
}
\label{fig::views}
\end{figure}

In this paper, we focus on a specific technique called ray-tracing volume rendering \citep{Levoy:1988:DSV:44650.44652}. The ray-tracing technique ``shoots'' rays through the cube. The final image is constructed by assigning pixel values as a function of voxel values sampled along the rays. Figure \ref{fig::views} shows different points-of-view of a volume rendered spectral cube of the barred spiral galaxy NGC 2903. The data is taken from The HI Nearby Galaxy Survey \citep[THINGS;][]{Walter2008AJ}. The different views show the result of both parallel and perspective projection (see Appendix \ref{app::projections} for more details). 

A transfer function is used to combine voxels encountered by each ray, where the result of this function is used to map the colour and intensity (or transparency) level of each pixel in the final image -- often using a {\em colour map}. A colour map is a lookup table used to associate a given scalar to a colour. To interpret the colour map, a colour bar is commonly displayed alongside the main visualisation (Figure \ref{fig::views}c.). The transparency of a pixel can vary from being fully opaque (no transparency) to being fully transparent. 

Two ray-tracing colouring technique are commonly used (see Section \ref{sec::transfer_functions} for examples). The first technique only combines a scalar value for each voxel encountered by the ray, and then either renders the resulting scalar as a greyscale, or optionally maps the scalar to a colour for display. The second technique combines full colour information as the ray is being traced, where the data is mapped to colour before combination.

As an early example, the {\tt Karma} software suite \citep{Gooch1996ASPC..101...80G} included 3D visualisation through texture-based volume rendering \citep{gooch1995astronomers}. Today, several options are available to visualise spectral cube using volume rendering. 

A number of recent solutions have considered using general purpose 3D visualisation software, including 3D Slicer \citep{Borkin2005astro.ph..6604B, PunzoEtAl2015A&C, Punzo201745}, Blender \citep{Kent2013PASP..125..731K, Taylor2015A&C....13...67T, Naiman2016A&C....15...50N, Garate2016arXiv161106965G}, Drishti \citep{Limaye2012SPIE.8506E..0XL, Potter0004-637X-794-2-174}, Houdini \citep{Naiman2017arXiv170101730N}, and Unity \citep{Ferrand2016arXiv160708874F}. This path reduces the need for astronomers to develop and maintain custom software. 

Another approach relies on developing custom software using visualisation libraries like {\tt VTK} \citep{Hanwell2015SoftX...1....9H}, {\tt S2PLOT} \citep[][]{Barnes2006PASA...23...82B}, 
or {\tt VisPy} \citep{campagnola:hal-01208191}. This provides control over the final product, but possibly at the cost of development time and maintenance. 

Finally, for cases where spectral cube data is larger than local memory (e.g. terabyte-scale spectral cube), distributed volume rendering frameworks running on supercomputers have also been considered \citep{
Hassan2013MNRAS.429.2442H}.

\subsection{Classic colouring method for volume rendered spectral cube data}
\label{sec::moments}
In a discussion about scientific visualisation challenges related to making discoveries in low signal-to-noise data, \citet{HassanFluke2011PASA} noted that the advanced use of colour to enhance comprehension has received little attention. Notable exceptions to this are \citet{Rector2005ca05.conf..194R}, \citet{Rector1538-3881-133-2-598}, and \citet{English2017IJMPD..2630010E} discussing the use of colour for presentation-quality astronomical images, and \citet{Ferrand2016arXiv160708874F} discussing use of colour for volume rendering visualisation of spectral cube data. 

In the context of volume rendering, the advanced use of colour refers to the development and usage of meaningful ways to map the scalar value and other properties of each volumetric element (or voxel) -- with an aim to distinguish the signal from the noise. 

An example of a transfer function used to visualise spectral cube data \citep[][]{Hassan2011100} is the Maximum Intensity Projection \citep[MIP; ][]{Wallis41482}. The maximum voxel for each traced ray is given by: 
\begin{equation}
\displaystyle{c_i = \left\{ \begin{array}{rl}
 v_i & \text{if } v_i \geq c_{i-1} \\
 c_{i-1} &\text{otherwise,}
  \end{array} \right.}
\label{eq::MIP}
\end{equation} 
where $v_i$ is the intensity of the $i$th voxel, $c_{i-1}$ is the solution based on the previously encountered voxels, and $c_i$ is the maximum intensity value at step $i$; $i$ increases as the ray is being traced. Similarly, the Accumulated Voxel Intensity Projection (AVIP) 
was used by \citet[][]{gooch1995astronomers} and \citet{Oosterloo1995PASA...12..215O} to create a transfer function that solves the radiative transfer equation for each traced ray: 
\begin{equation}
\displaystyle{c_i = k_iv_i + (1-k_i)c_{i-1}.}
\label{eq::AVIP}
\end{equation} 
Here $k_i$ and $v_i$ are the transparency level and intensity of voxel $i$ respectively; $c_{i-1}$ is the solution based on the previously encountered voxels; and $c_i$ is the sum after this voxel is added. Using $k\propto v^\alpha$, where $\alpha$ is a weighting parameter, they provided a way to modify the transparency level to highlight or hide features in the data as required. 

To the best of our knowledge, all volume rendered 3D visualisations in astronomy relied on the direct use of the scalar output of the transfer functions to map the pixel colour --- while transparency is either not considered (e.g. all colours are opaque), set by the scalar, or set using a manually defined function to mask certain ranges of the transfer function domain. We note that other software have moved beyond the scalar ray-tracing technique where a one dimensional colour map lookup is used. For example, Drishti offers 2D colour map lookup, and volumetric effects in computer games are now routinely doing vector ray-tracing.

In the form presented above, MIP and AVIP are similar in concept to the zeroth moment map. For example, computing the zeroth moment map of a spectral cube consists of evaluating the integrated flux for each spectrum:
\begin{equation}
\displaystyle{M0 = \int v_i \Delta i \approx \Delta i \sum v_i,}
\label{eq::mom0}
\end{equation}
where $\Delta i$ is the bin width of the spectrum, $v_i$ is the $i$th spectral channel, and M0 is the resulting zeroth moment scalar. Computing a moment map is equivalent to the ray-tracing method, tracing rays through the spectral axis using a parallel projection.

Directly converting MIP or AVIP to colour and transparency will provide information about the integrated flux only. Information about the distribution of the flux along the spectral dimension can only be accessed through visualisation from several angles --- and this is precisely why 3D visualisation, including stereoscopic methods, has been used. To access this information in 2D, the first moment map provides a view of 
the per-voxel intensity-weighted velocity function: the velocity field.  For each spectrum in the spectral cube:
\begin{equation}
\displaystyle{M1 = \frac{\int iv_i \Delta i}{\int v_i \Delta i} \approx \frac{\sum iv_i }{\sum v_i}.}
\label{eq::mom1}
\end{equation}
A variant of this algorithm --- behaving similarly to that of MIP in relation to $M0$ --- is to compute the $\argmax$ function to retrieve the index of the maximum value in the spectrum. 

As moment map methods compute Equations \ref{eq::mom0} or \ref{eq::mom1} for each line of sight of the spectral cube, a serial implementation of the algorithm is inefficient. As each spectrum can be computed independently, the algorithm qualifies as what is generally called an ``embarrassingly parallel problem'', and is well suited for massively parallel hardware and the Stream Programming Model.

\subsection{GPU shaders and the Stream Programming Model}
\label{sec::stream_programming}

The Stream Programming Model consists of structuring applications in a way that allows high efficiency in computation and communication \citep{Kapasi:2003:PSP:939824.939856}. It is the main programming model for GPUs. In the stream programming model, a {\em stream} is an ordered set of data of the same data type. Data type can be simple (e.g. integers or floats) or complex (e.g. points, triangles, transformation matrices). Computation on streams is performed by a series of {\em kernels}. A kernel acts on entire streams, taking one or more streams as inputs and producing one or more streams as outputs. Applications are constructed by chaining multiple kernels together.

Volume rendering is a good match for the stream model, as the graphics pipeline is structured as stages of computation connected by data flow between the stages, similar to the stream and kernel abstractions. 

\begin{figure}
\includegraphics[width=8.3cm]{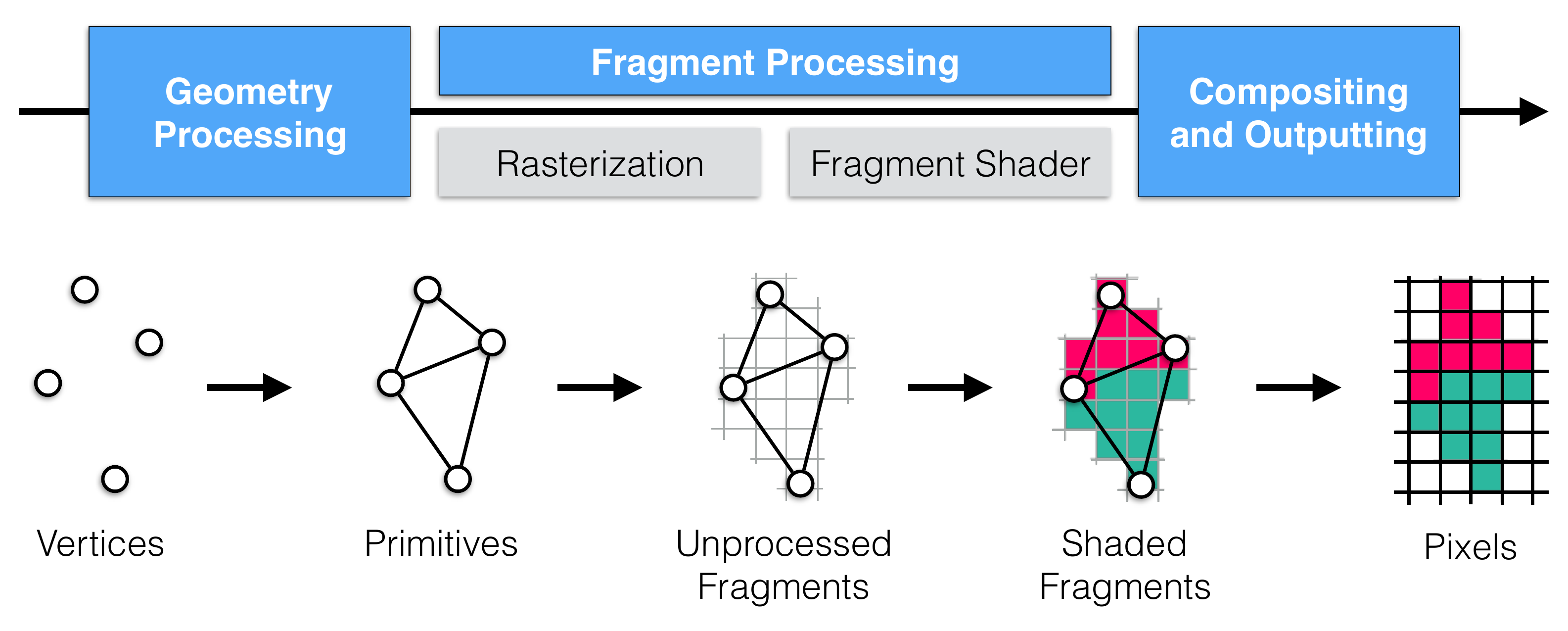}
\caption{The graphics pipeline as a stream programming model [adapted from \citet{Engel:2004:RVG:1103900.1103929}]. Three stages of the pipeline are highlighted (blue boxes): (1) geometry processing; (2) fragment processing; and (3) compositing and outputting to buffer. In addition, the two steps of fragment processing are shown (grey boxes): the rasterization step discretizes primitives into fragments; the fragment shader sets the colour and transparency level for each fragment based on a user-defined program. The input and output of each step is shown below the boxes.}
\label{fig::graphics_pipeline}
\end{figure}

The current graphics pipeline allows programmers to define a number of stages manually through kernel programs called {\em shaders} (Figure \ref{fig::graphics_pipeline}). Thus, for spectral cube volume rendering, we can take advantage of the flexibility offered by the {\em fragment shader} to compute the transfer function for each ray in parallel on the GPU.

Shader programs are generally written using a shading language [e.g. OpenGL Shading Language\footnote{\url{https://www.khronos.org/opengl/wiki/OpenGL_Shading_Language}} (GLSL) and High-Level Shading Language\footnote{\url{https://msdn.microsoft.com/en-us/library/windows/desktop/bb509561(v=vs.85).aspx}} (HLSL)]. Shading languages provide different data types which can be used to considerably simplify the generation of the final image. For example, in addition to {\tt float} and {\tt int} types, GLSL provides many additional types, including vector types ({\tt vec2}, {\tt vec3}, and {\tt vec4}, each comprising two, three and four floating point value respectively). Vector types provide a simple way to encode visual information. 

By convention in shading langages, values of a vector can be accessed directly using spatial (x,y,z,w) or colour space [red (r), green (g), blue (b), alpha (a)] variables (e.g. $MyVec4Vector.x$ or $MyVec4Vector.r$ will return the first value of the $MyVec4Vector$ vector as a floating point number, and $MyVec4Vector.xya$ would return a {\tt vec3} array composed of the first, second and fourth values of the vector. We use this convention in Section \ref{sec::transfer_functions}. These key features of the modern graphics pipeline now allow computation of transfer functions at run-time in a way that was not available in the past. 

\section{Transfer functions and fragment shaders}
\label{sec::transfer_functions}
In this section, we discuss three colouring methods that compute and render different attributes of the data: integrated flux (Section \ref{sec::m0}), velocity field (Section \ref{sec::m1}), and 3D distribution (Section \ref{sec::rgb}). 

In the following, we will limit our attention to the MIP and AVIP methods described previously as the starting point to assign the colour and transparency of pixels in the final image. For the sake of clarity, we rewrite the MIP and AVIP equations as part of the ray-tracing algorithm to highlight how we proceed using the fragment shader. The main tasks of the ray-tracing algorithm, assuming a ray visits all voxels in a front-to-back order, are shown in Algorithm \ref{algo::ray-tracing}. For all algorithms, we use the following writing convention:
\begin{itemize}
\item a variable (of type {\tt int}, {\tt float}, {\tt vec4}, {\tt tex3D}, ...) is represented by a name in italic (e.g. {\em cube});
\item a function is represented by a name in regular text, and its parameters are included inside parenthesis (e.g. myFunction({\em variable}));
\item a value at location {\em loc} inside an array (e.g. the 3D array representing the spectral cube of type {\tt tex3D}) is accessed using square brackets (e.g. {\em vector}[{\em loc}]);
\item values of a vector ({\tt vec2}, {\tt vec3}, or {\tt vec4}) can be accessed using a point, as described in Section \ref{sec::stream_programming} (e.g. {\em vector.rgb}). 
\end{itemize}
	
\begin{algorithm}
  \caption{Outline of the front-to-back ray-tracing algorithm}
  \begin{algorithmic}[1]
    \Require{$cube$: the 3D array; $step$: vector of length 3 representing ray increment; $loc$: vector of length 3 representing a location in 3D space; $N$: number of steps to take along the ray.}
    \Statex
    \Function{rayTracing}{}
      \Let{$vars$}{initialise($vars$)}
      \Let{$loc$}{$startLoc$}
      \For{$i \gets 1 \textrm{ to } N$}
      	 \Let{$val$}{$cube[loc]$}
          \Let{$val$}{smooth($vars$)} \Comment{Optionally smooth value}
          \Let{$val$}{filter($vars$)} \Comment{Optionally filter or mask value}
          \Let{$vars$}{transferFunction($vars$)}
          \Let{$loc$}{$loc$ + $step$}
      \EndFor
      \Let{$fragment$}{setFragment($vars$)}
      \State \Return{$fragment$}
    \EndFunction
  \end{algorithmic}
  \label{algo::ray-tracing}
\end{algorithm}

In Algorithm \ref{algo::ray-tracing}, {\tt initialise()}, {\tt transferFunction()} and {\tt setFragment()} are functions that will vary for each colouring method. Each variation of {\tt transferFunction()} and {\tt setFragment()} is outlined in Sections \ref{sec::m0}, \ref{sec::m1}, and \ref{sec::rgb}; {\tt initialise()} simply prepares the variables required to compute the transfer function. 

To simplify the pseudo code of Algorithm \ref{algo::ray-tracing}, we use $vars$ to represent any variables a function may require as parameter; for example, some instances of {\tt transferFunction()} require the variables $val$ and $loc$, while other instances may not. In all cases, the returned $fragment$ will be a vector of length 4, containing the RGBA information of a given fragment (pixel in the final image). 
Variations of the functions {\tt smooth()} and {\tt filter()} are outlined in Section \ref{sec::Filters}. On a modern graphics card, Algorithm \ref{algo::ray-tracing} will be executed in parallel for each fragment at run-time. 

\subsection{Zeroth moment-inspired transfer functions}
\label{sec::m0}
In Section \ref{sec::moments}, we described how to compute MIP (Equation \ref{eq::MIP}) and AVIP (Equation \ref{eq::AVIP}), and how the resulting scalar is directly used to set the colour, and potentially the transparency level, of a given pixel in the final image. 

Algorithms \ref{algo::MIP0} and \ref{algo::AVIP0} show the {\tt transferFunction()} and {\tt setFragment()} functions for MIP and AVIP respectively. We modify equation \ref{eq::AVIP} to follow the front-to-back process of Algorithm \ref{algo::ray-tracing} with an arbitrary weighting factor. 

For MIP, the scalar is used to set the colour using a colour map, and the transparency level can optionally be set using user defined criteria. For example, one can simply set all colours to be fully opaque as in Figure \ref{fig::views}. For AVIP, the colour is computed as the weighted average value along the ray, where the weighting parameter is used to set the opacity. In the following, we refer to this direct mapping as MIP$_0$ and AVIP$_0$, as they produce similar visualisations as the zeroth moment map (Figure \ref{fig::views}).

\begin{algorithm}
  \caption{MIP$_0$}
  \begin{algorithmic}[1]
    \Require{$maxval$: initialized to the smallest possible value; colourMap($scalar$): a function that maps a scalar to colour and transparency (RGBA).}
    \Statex
    \Function{transferFunction}{$val, maxval$}
	\If{$val \geq maxval$}
		\Let{$maxval$}{$val$}
        \EndIf
      \State \Return{$maxval$}
    \EndFunction
    \\
    \Function{setFragment}{$maxval$}
	\Let{$fragment$}{colourMap($maxval$)}
      \State \Return{$fragment$}
    \EndFunction
  \end{algorithmic}
  \label{algo::MIP0}
\end{algorithm}

\begin{algorithm}
  \caption{AVIP$_0$}
  \begin{algorithmic}[1]
    \Require{$tempFrag$: a temporary fragment (RGBA); $k$: an arbitrary weighting factor, $minVal$: arbitrary small value greater than 0. colourMap($scalar$): a function that maps a scalar to colour and transparency (RGBA), max(val1, val2): a function returning the largest of two scalar.}
    \Statex
    \Function{transferFunction}{$val, tempFrag, k$}
	\Let{$colour$}{colourMap($val$)}
	\Let{$\alpha_1$}{$tempFrag.a$}
	\Let{$\alpha_2$}{$val \times k \times (1-\alpha_1)$}
	\Let{$\alpha$}{max($\alpha_1 + \alpha_2, minVal)$}
	\Let{$tempFrag$}{$tempFrag \times \frac{\alpha_1}{\alpha}+colour \times \frac{\alpha_2}{\alpha}$}
	\Let{$tempFrag.a$}{$\alpha$}
	\If{$\alpha > 0.99$}
		\Let{$i$}{$N$}
        \EndIf
      \State \Return{$tempFrag$}
    \EndFunction
    \\
    \Function{setFragment}{$maxval$}
	\Let{$fragment$}{$tempFrag$}
      \State \Return{$fragment$}
    \EndFunction
  \end{algorithmic}
  \label{algo::AVIP0}
\end{algorithm}

\subsection{First moment-inspired transfer functions}
\label{sec::m1}

\begin{figure*}
\includegraphics[width=17.5cm]{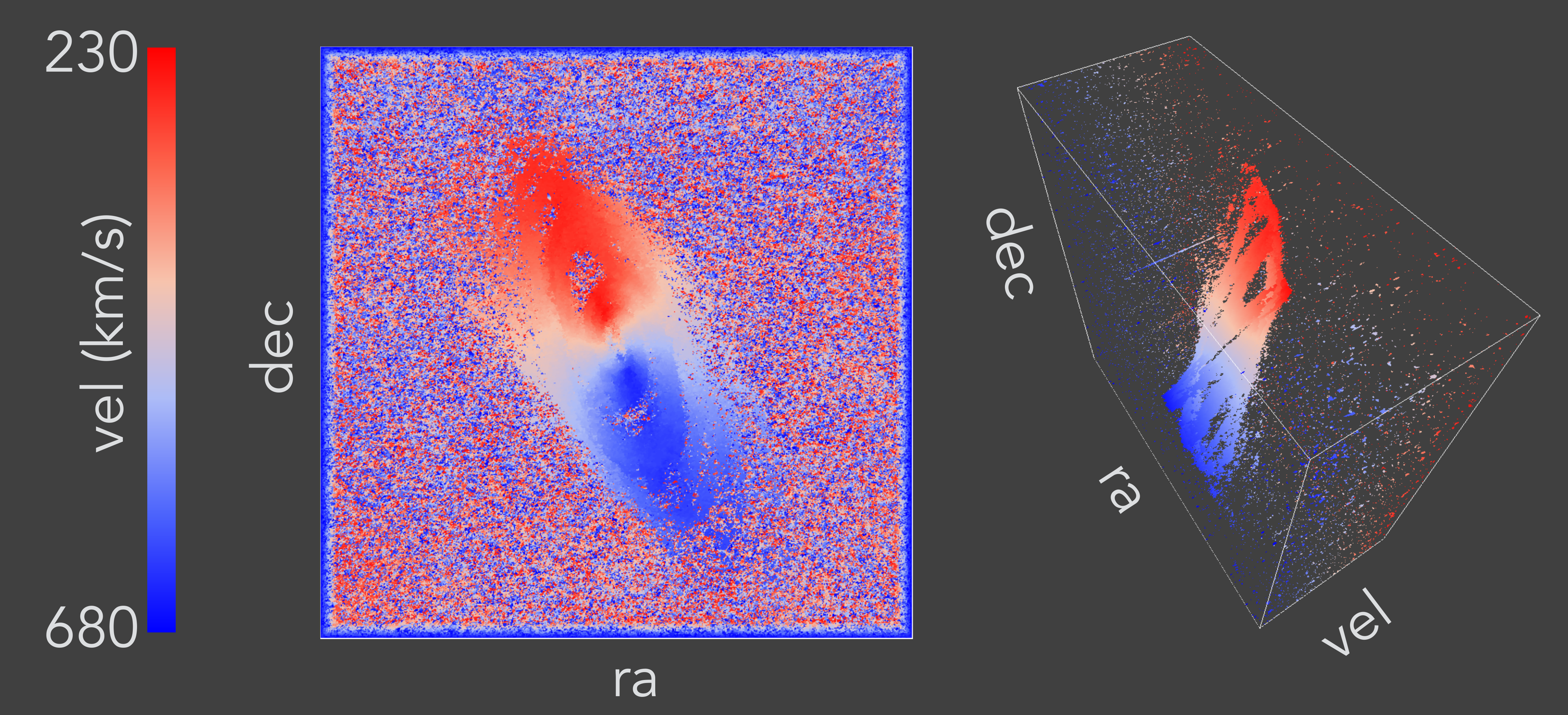}\\
\caption{Volume rendering of NGC 2903 using MIP$_{0}$. Right panel shows the spatial view, while left panel shows a blend of spatial and spectral views. Details of the visualisation parameters are described in Section \ref{sec::discussion}.}
\label{fig::mip-no-alpha}
\end{figure*}

We can modify Algorithms \ref{algo::MIP0} and \ref{algo::AVIP0} to produce visualisations inspired by the first moment map (Figure \ref{fig::mip-no-alpha}). To do so, instead of mapping the colour from the colour map using the voxel intensity, we will use the velocity or redshift of the voxel(s) of interest. The voxel intensity can then be used to set the transparency level as required. The new algorithms MIP$_1$ and AVIP$_1$ are shown in Algorithms \ref{algo::MIP1} and \ref{algo::AVIP1}, where respective variations from Algorithms \ref{algo::MIP0} (MIP$_{0}$) and \ref{algo::AVIP0} (AVIP$_{0}$) are highlighted in bold. 

\begin{algorithm}
  \caption{MIP$_1$}
  \begin{algorithmic}[1]
    \Require{$maxval$: initialized to the smallest possible value; $maxloc$: coordinate of maximum voxel initialised to 0; colourMap($scalar$): a function that maps a scalar to colour and transparency (RGBA).}
    \Statex
    \Function{transferFunction}{$val, loc, maxval, maxloc$}
	\If{$val \geq maxval$}
		\Let{$maxval$}{$val$}
		\Let{$maxloc$}{$loc$}
        \EndIf
      \State \Return{$maxval, maxloc$}
    \EndFunction
    \\
    \Function{setFragment}{maxval, maxloc}
	\Let{{\bf \em fragment}}{{\bf colourMap({\em maxloc.z})}}
	\Let{{\bf \em fragment.a}}{{\bf \em maxval}}
      \State \Return{$fragment$}
    \EndFunction
  \end{algorithmic}
  \label{algo::MIP1}
\end{algorithm}

\begin{algorithm}
  \caption{AVIP$_1$}
  \begin{algorithmic}[1]
    \Require{$tempFrag$: a temporary fragment (RGBA); $k$: an arbitrary weighting factor, $minVal$: arbitrary small value greater than 0. colourMap($scalar$): a function that maps a scalar to colour and transparency (RGBA), max(val1, val2): a function returning the largest of two scalar.}
    \Statex
    \Function{transferFunction}{$val, loc, tempFrag, k$}
	\Let{{\bf \em colour}}{{\bf colourMap(\em loc.z)}}
	\Let{$\alpha_1$}{$tempFrag.a$}
	\Let{$\alpha_2$}{$val \times k \times (1-\alpha_1)$}
	\Let{$\alpha$}{max($\alpha_1 + \alpha_2, minVal)$}
	\Let{$tempFrag$}{$tempFrag \times \frac{\alpha_1}{\alpha}+colour \times \frac{\alpha_2}{\alpha}$}
	\Let{$tempFrag.a$}{$\alpha$}
	\If{$\alpha > 0.99$}
		\Let{$i$}{$N$}
        \EndIf
      \State \Return{$tempFrag$}
    \EndFunction
    \\
    \Function{setFragment}{$tempFrag$}
	\Let{$fragment$}{$tempFrag$}
      \State \Return{$fragment$}
    \EndFunction
  \end{algorithmic}
  \label{algo::AVIP1}
\end{algorithm}

This type of transfer function can inform about two types of information at once: the velocity of maximal or integrated emission, and the relative intensity of emission via the transparency level. Along with MIP$_1$ and AVIP$_1$, the colour bar should provide information about the velocity range being visualised. It is an addition from the zeroth moment-inspired mode where this information was only available through animation. From certain viewing positions like in Figure \ref{fig::views}d ({\em what is the `dec' coordinate of the signal?}) and Figure \ref{fig::views}e ({\em what is the `ra' coordinate of the signal?}), it can be difficult to evaluate the spatial coordinates of a pixel in the final image using MIP$_1$ and AVIP$_1$. In the next section, we introduce a qualitative transfer function that can help to provide information about all three dimensions.

\subsection{RGB cube transfer functions}
\label{sec::rgb}
\begin{figure*}
\includegraphics[width=17.5cm]{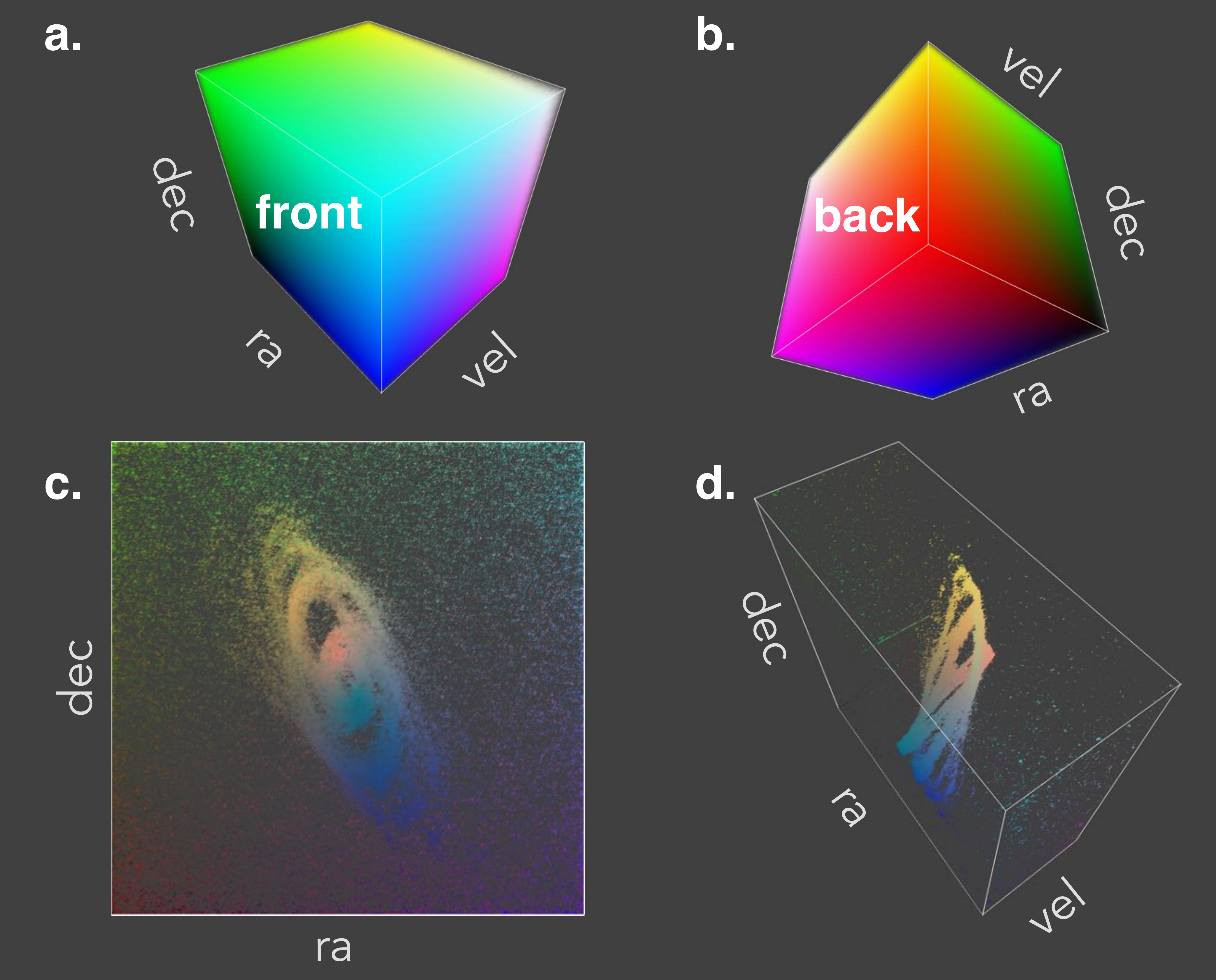}\\
\caption{RGB cube transfer function. The RGB colour space is shown in relation to spatial and velocity axes: {\bf a.} and {\bf b.} colours position in relation to the front and the back faces of the same cube respectively, and mapping to ra, dec and vel; {\bf c.} the spatial view of NGC 2903 showing ra and dec; {\bf d.} a blend of spatial and spectral views showing ra, dec and vel.}
\label{fig::rgb_algo}
\end{figure*}

We can generalise MIP$_1$ and AVIP$_1$ by mapping all three dimensions to colour. As we have access to the RGB colour space, which is a 3D space, we can use it to provide a unique colour to every 3D coordinate of the spectral cube (Figure \ref{fig::rgb_algo}). 

To do so, we modify the previous algorithms to create MIP$_{RGB}$ and AVIP$_{RGB}$ as shown in Algorithms \ref{algo::RGB-MIP} and \ref{algo::RGB-AVIP}. Here again, respective variations from Algorithms \ref{algo::MIP0} (MIP$_{0}$) and \ref{algo::AVIP0} (AVIP$_{0}$) are highlighted in bold. The RGB cube transfer function will provide information about both spatial and spectral coordinates. It is, however, a purely qualitative visualisation. In this mode of operation there is no colour bar involved. Instead, a colour reference cube may provide a handy visual link between the RGB space and the three coordinate axes. 

\begin{algorithm}
  \caption{MIP$_{RGB}$}
  \begin{algorithmic}[1]
    \Require{$cube$: the 3D array; $maxval$: initialized to the smallest possible value; $maxloc$: coordinate of maximum voxel initialised to 0; colourMap($scalar$): a function that maps a scalar to colour and transparency (RGBA).}
    \Statex
    \Function{transferFunction}{$cube, val, loc, maxval, maxloc$}
	\If{$cube[loc] \geq maxval$}
		\Let{$maxval$}{$val$}
		\Let{$maxloc$}{$loc$}
        \EndIf
      \State \Return{$maxval, maxloc$}
    \EndFunction
    \\
    \Function{setFragment}{maxval, maxloc}
	\Let{{\bf \em fragment.r}}{{\bf \em maxloc.z}}
	\Let{{\bf \em fragment.g}}{{\bf \em maxloc.y}}
	\Let{{\bf \em fragment.b}}{{\bf \em maxloc.x}}
	\Let{{\bf \em fragment.a}}{{\bf \em maxval}}
      \State \Return{$fragment$}
    \EndFunction
  \end{algorithmic}
  \label{algo::RGB-MIP}
\end{algorithm}

\begin{algorithm}
  \caption{AVIP$_{RGB}$}
  \begin{algorithmic}[1]
    \Require{$tempFrag$: a temporary fragment (RGBA); $k$: an arbitrary weighting factor, $minVal$: arbitrary small value greater than 0. colourMap($scalar$): a function that maps a scalar to colour and transparency (RGBA), max(val1, val2): a function returning the largest of two scalar.}
    \Statex
    \Function{transferFunction}{$loc, val, tempFrag, k$}
	\Let{{\bf \em colour.r}}{{\bf \em loc.z}}
	\Let{{\bf \em colour.g}}{{\bf \em loc.y}}
	\Let{{\bf \em colour.b}}{{\bf \em loc.x}}
	\Let{$\alpha_1$}{$tempFrag.a$}
	\Let{$\alpha_2$}{$val \times k \times (1-\alpha_1)$}
	\Let{$\alpha$}{max($\alpha_1 + \alpha_2, minVal)$}
	\Let{$tempFrag$}{$tempFrag \times \frac{\alpha_1}{\alpha}+colour \times \frac{\alpha_2}{\alpha}$}
	\Let{$tempFrag.a$}{$\alpha$}
	\If{$\alpha > 0.99$}
		\Let{$i$}{$N$}
        \EndIf
      \State \Return{$tempFrag$}
    \EndFunction
    \\
    \Function{setFragment}{$tempFrag$}
	\Let{$fragment$}{$tempFrag$}
      \State \Return{$fragment$}
    \EndFunction
  \end{algorithmic}
  \label{algo::RGB-AVIP}
\end{algorithm}

\section{Filtering and fragment shaders}
\label{sec::Filters}
When performing a visual exploration of spectral cube data, we often have to deal with noisy data, where sources are very close to the noise level. In general, data needs to be pre-processed before interesting features become apparent during visualisation. For low signal-to-noise data, filtering techniques have been shown to enhance manual data inspection \citep{Oosterloo1996VA.....40..571O, Punzo2016A&C....17..163P}. Such techniques are commonplace in automated segmentation methodologies like source finding and source mask generation \citep{Whiting2012MNRAS.421.3242W, Serra2015MNRAS.448.1922S}, but have rarely been integrated with visualisation as they are often too compute-intensive. 

In this section, we present fragment shader kernels of four filtering techniques commonly used for data with signal extended over many pixels and small spatial intensity derivative; namely (1) {\em box smoothing}, (2) {\em Gaussian smoothing}, (3) {\em intensity clipping},  and (4) {\em intensity domain scaling}. The aim of this section is to highlight how such techniques can be incorporated in the graphics pipeline to interactively improve the visualization output. We evaluate their algorithmic complexity to estimate their effect on the rate at which new frames can be rendered (frame rate). Presenting thorough use cases and analyses of these techniques is beyond the scope of this paper. Instead, we refer the reader interested in a review of image processing techniques to \citet{Buadesdoi:10.1137/040616024} and \citet{goyal2012comprehensive}, and to \citet{Punzo2016A&C....17..163P} for application to HI spectral cube data. 

As shown in the previous section, we perform the ray-tracing algorithm within the fragment shader. In this context, the spectral cube data is loaded in GPU memory ({\em in-situ} visualisation) using a 3D array (or more precisely a 3D texture).  This introduces a limit on the number of voxels that can be volume rendered.  For a typical commodity GPU video random access memory of 2 GB, this is sufficient for a $1024^3$-voxel spectral cube.

With this in-core access to the data within GPU memory, we can further exploit the interactivity of shaders to perform filtering techniques at run-time. This is achieved by integrating filtering kernels as part of the graphics pipeline. These kernels can be optionally used as indicated at lines 6 and 7 of Algorithm \ref{algo::ray-tracing}. In the context of Algorithm \ref{algo::ray-tracing}, the function {\em smooth} (line 6) refers to techniques (1) and (2), and {\em filter} (line 7, dynamic histogram manipulation) refers to techniques (3) and (4).

\subsection{1D box smoothing}
\label{sec::box_filter}
Box smoothing \citep{MCDONNELL198165} is a convolution kernel used to smooth the data. The algorithm proceeds by replacing the value of a voxel with the average of neighbouring voxels inside a box. In the present context, we consider a 1D box smoothing computed along the ray direction. The size of the box is an odd number, so as to compute the average of the voxel itself with a symetrical number of neighbours. This convolution filter is a simple method that smooths data and reduces noise in the final image. The box smoothing algorithm is shown as a fragment shader kernel in Algorithm \ref{algo::box}.  

\begin{algorithm}
  \caption{box smoothing}
  \begin{algorithmic}[1]
    \Require{$cube$: the 3D array; $loc$: current voxel location; $filterArm$: number of neighbouring voxels to visit on each side of the original voxel; $filterCoe\hspace{-0.1em}f\hspace{-0.2em}f$: box size$^{-1}$ = $(2\times filterArm+1)^{-1}$.}
    \Statex
    \Function{smooth}{$cube, loc, filterArm, filterCoe\hspace{-0.1em}f\hspace{-0.2em}f$}
	\Let{$val$}{$cube[loc]$}
	\For{$i \gets 1 \textrm{ to } filterArm$}
		\Let{$val$}{$val + cube[loc+i]$}
		\Let{$val$}{$val + cube[loc-i]$}
	\EndFor
	\Let{$val$}{$val \times filterCoe\hspace{-0.1em}f\hspace{-0.2em}f$}
      \State \Return{$val$}
    \EndFunction
  \end{algorithmic}
  \label{algo::box}
\end{algorithm}

The box smoothing algorithm as defined here has a complexity of $O(N)$, where $N$ is the number of voxels in the box (voxel lookup)\footnote{To reduce the number of texture fetch $N$, alternative algorithms that keep previously visited voxel values in memory could be considered. This however comes at the cost of being more memory hungry. Our tests showed a minor slow down in frame rate using such a technique.}. In conjunction with ray-tracing, the total algorithm becomes $O(NM)$, where $M$ is the number of voxels visited by a ray, and $N$ the number of voxels visited by the box smoothing.

\subsection{3D Gaussian smoothing}
\label{sec::Gaussian_filter}
Our next smoothing kernel is Gaussian smoothing, which consists of convolving a pixel with a 3D Gaussian distribution. Hence, instead of giving each neighbour the same weight, as does the box smoothing, the Gaussian smoothing's weights follow a distribution such that: 
\begin{equation}
G(x,y,z) = A \exp^{-\left(\frac{(x-x_0)^2)}{2\sigma_x^2} + 
\frac{(y-y_0)^2)}{2\sigma_y^2} + \frac{(z-z_0)^2)}{2\sigma_z^2}\right)},
\label{eq::gauss}
\end{equation} 
where $\sigma_x$, $\sigma_y$, and $\sigma_z$ are linked to the full width at half maximum (FWHM) of the peak, determining the amount of smoothing:
\begin{equation}	
FWHM_i = 2\sqrt{2\ln (2)} \sigma_i.
\label{eq::fwhm}
\end{equation} 
For simplicity, we only consider isotropic Gaussian kernels; however, a similar solution to the one presented below could be suitable for an anisotropic kernel. A fast algorithm that gives a good approximation to the  Gaussian distribution consists of a discrete sampling of coefficients obtained with cascaded convolution of a kernel filter composed of [1,1] \citep{Crowley02fastcomputation}. The coefficients for the {\em n}th filter in the series, $b_n(m)$, are defined by:
\begin{equation}
b_n(m) = [1, 1]^{*n},
\end{equation}
where the exponent $*n$ denotes $n$ autoconvolutions. The set of coefficients is  the binomial series, which can be pre-computed using Pascal's triangle; the series provides its best approximation for a Gaussian of finite size \citep{Crowley02fastcomputation}. In particular, $b_4(n)$, $b_8(n)$, and $b_{12}(n)$ are cases of interest for isotropic kernels, which are 5-tap, 9-tap, and 13-tap Gaussian smoothings respectively, named in reference to the total number of voxel lookup required for each filtered voxel\footnote{We base the calculation of the coefficient and determination of lookup coordinate for the binomial filter on the formula presented by Daniel R\'{a}kos in a \href{http://rastergrid.com/blog/2010/09/efficient-Gaussian-blur-with-linear-sampling/}{blog post} (last accessed 9 February 2017). In particular, R\'{a}kos proposes that the calculation for the weight and offset make use of the bilinear texture filtering offered by the GPU. Doing so, it is possible to get information about two voxels at once if we don't fetch voxel at its center positions, reducing the overall number of texture fetch required for a given kernel size. We refer the reader to the article for further clarifications.}. 

To obtain a 3D kernel along the data axes, one simply needs to successively apply the 1D convolution on each axis. We show the case of the 9-tap filter algorithm as a fragment shader kernel in Algorithm \ref{algo::9-tap}. 

\begin{algorithm}
  \caption{9-tap Gaussian smoothing}
  \begin{algorithmic}[1]
    \Require{$cube$: the 3D array; $oVal$: original voxel value (before convolution); $loc$: current voxel location; ${\mathit resolution}$: spectral cube resolution (dimensions); ${\mathit direction}$: rotation matrix to select axis onto which lookups are done.}
    \Statex
    \Function{smooth}{$cube, oVal, loc, {\mathit resolution}, {\mathit direction}$}
	\Let{$val$}{$(0,0,0,0)$}
	\Let{$o\hspace{-0.1em}f\hspace{-0.2em}f_1$}{$vec3(1.3846153846)\times {\mathit direction}$}
	\Let{$o\hspace{-0.1em}f\hspace{-0.2em}f_2$}{$vec3(3.2307692308)\times {\mathit direction}$}
	\Let{$coe\hspace{-0.1em}f\hspace{-0.2em}f_0$}{$0.2270270270$}
	\Let{$coe\hspace{-0.1em}f\hspace{-0.2em}f_1$}{$0.3162162162$}
	\Let{$coe\hspace{-0.1em}f\hspace{-0.2em}f_2$}{$0.0702702703$}
	\Let{$val$}{$val+oVal\times {\mathit coe\hspace{-0.1em}f\hspace{-0.2em}f_0}$}
	\Let{$tempVal$}{$cube[loc+\frac{o\hspace{-0.1em}f\hspace{-0.2em}f_1}{{\mathit resolution}}]$}
	\Let{$val$}{$val+tempVal\times coe\hspace{-0.1em}f\hspace{-0.2em}f_1$}
	\Let{$tempVal$}{$cube[loc-\frac{o\hspace{-0.1em}f\hspace{-0.2em}f_1}{resolution}]$}
	\Let{$val$}{$val+tempVal\times coe\hspace{-0.1em}f\hspace{-0.2em}f_1$}
	\Let{$tempVal$}{$cube[loc+\frac{off_2}{resolution}]$}
	\Let{$val$}{$val+tempVal\times coe\hspace{-0.1em}f\hspace{-0.2em}f_2$}
	\Let{$tempVal$}{$cube[loc-\frac{o\hspace{-0.1em}f\hspace{-0.2em}f_2}{resolution}]$}
	\Let{$val$}{$val+tempVal\times coe\hspace{-0.1em}f\hspace{-0.2em}f_2$}
      \State \Return{$val$}
    \EndFunction
  \end{algorithmic}
  \label{algo::9-tap}
\end{algorithm}

The tap-filter algorithm has a complexity of $O(N)$, where $N$ is the number of lookups. In conjunction with ray-tracing, the total algorithm becomes $O(NMD)$, where $M$ is the number of voxels visited by a ray, $N$ the number of lookups, and $D$ the dimensionality of the convolution kernel (1-D, 2-D, 3-D, ..., N-D). The kernel size (e.g. 9-tap) will have an effect on the rendering speed.  However, it is worth noting that in our example, M and D are small [e.g. $M\in(5, 9, 13)$, and $D=3$], and hence can be handled by modern graphics (depending of course also on the spectral cube size). 

\subsection{Intensity clipping}

Intensity clipping is a straight-forward technique consisting of setting a threshold beyond which values are discarded. The threshold value is set interactively at run-time. In this work, we consider a minimum and a maximum threshold value for interactive data exploration. The minimum threshold can be used, for example, to discard low intensity noise voxels, while the maximum threshold can be used to mask very intense voxels from sources like stars. The algorithm is shown as a fragment shader kernel in Algorithm \ref{algo::clipping}. The complexity of the intensity clipping is $O(1)$.

\begin{algorithm}
  \caption{Intensity clipping}
  \begin{algorithmic}[1]
    \Require{$minT\hspace{-0.1em}hreshold$: minimum threshold value; $maxT\hspace{-0.1em}hreshold$: maximum threshold value; $DAT\hspace{-0.1em}AMIN$: lower bound of data range.}
    \Statex
    \Function{filter}{$val, minT\hspace{-0.1em}hreshold, maxT\hspace{-0.1em}hreshold$}
	\If{$val < minT\hspace{-0.1em}hreshold$}
		\Let{$val$}{$DAT\hspace{-0.1em}AMIN$}
        \EndIf
        
        	\If{$val > maxT\hspace{-0.1em}hreshold$}
		\Let{$val$}{$DAT\hspace{-0.1em}AMIN$}
        \EndIf
      \State \Return{$val$}
    \EndFunction
  \end{algorithmic}
  \label{algo::clipping}
\end{algorithm}

\subsection{Intensity domain scaling}

Intensity domain scaling is a similar technique to intensity clipping, setting thresholds in the data intensity range. However, instead of simply discarding values while keeping the same colour map, the domain of intensity values is rescaled so that the colour map's minimum and maximum match this rescaled domain. This technique provides a way to ``zoom in'' on a section of an intensity range of interest. The algorithm is shown as a fragment shader kernel in Algorithm \ref{algo::scale}. The complexity of the intensity domain scaling is $O(1)$.

\begin{algorithm}
  \caption{Intensity domain scaling}
  \begin{algorithmic}[1]
    \Require{$minT\hspace{-0.1em}hreshold$: minimum threshold value; $maxT\hspace{-0.1em}hreshold$: maximum threshold value; insure that $minT\hspace{-0.1em}hreshold \leq maxT\hspace{-0.1em}hreshold$; $DAT\hspace{-0.1em}AMIN$: lower bound of data range.}
    \Statex
    \Function{filter}{$val, minT\hspace{-0.1em}hreshold, maxT\hspace{-0.1em}hreshold$}
    	\Let{$discardRatio$}{$1.0 / (maxT\hspace{-0.1em}hreshold - minT\hspace{-0.1em}hreshold)$}
	
	\If{$val > maxT\hspace{-0.1em}hreshold$}
		\Let{$val$}{$DAT\hspace{-0.1em}AMIN$}
	\ElsIf{val < minT\hspace{-0.1em}hreshold}
		\Let{$val$}{$DAT\hspace{-0.1em}AMIN$}
	\Else
		\Let{$val$}{$val-minT\hspace{-0.1em}hreshold$}
		\Let{$val$}{$val \times discardRatio$}
        \EndIf
	\State \Return{$val$}
    \EndFunction
  \end{algorithmic}
  \label{algo::scale}
\end{algorithm}

 \subsection{Computing an emission line ratio in 3D space}
 \label{app:3D-line-ratio}
 
We now show how a more complex task --- computing an emission line ratio between two data cubes --- can be achieved with real-time ray tracing.
 
Observations of molecular gas in galaxies (e.g. the Antennae cube) or planetary nebula are often performed for several molecular transitions resulting in several spectral cubes of the same object. Computing the ratio between these transitions provides useful insights on, for example, the excitation of the gas or its metallicity --- adding physical information to the rendering. Since this calculation is computed for each spectra of the cubes, it can also be computed through ray-tracing volume rendering. 

As a proof of concept, this Section presents a modification of our ray-tracing volume rendering algorithm (Algorithm \ref{algo::ray-tracing}) to compute the ratio between two emission lines. In practice, extra steps should be added to the algorithm (e.g. line fitting). The exact details of the method will vary based on the different use cases and scientific questions to be explored. Exploring all cases in detail goes beyond the scope of this paper. Here, we present two methods to compute the line ratio. The first method shares similarities with the traditional method of computing a line ratio, while the second method computes ratios on a voxel by voxel basis.  

To compute line ratios, the two spectral cubes --- sub cubes of a unique object observed at different wavelengths --- are set to a unified grid (e.g. same number of voxels for all dimensions), and are loaded in two separate 3D textures on the GPU, filling the same 3D space. At each step taken by the ray, both spectral cubes can be sampled using the same grid location $loc$.

{\bf Method 1.} 
A metric (e.g. MIP or AVIP) is computed for both spectral cubes independently. After all voxels in the path of the ray have been visited, the ratio between the two results is computed. Computing the ratio in this manner (e.g. with MIP) will provide information about the relation between the maximum flux from both emission lines. The resulting ratio is used to set the colour and/or transparency of the pixel via the colour map --- making sure the ratio is normalized to the range accepted by the visualisation library before setting the colour (i.e. [0, 1]). The method is shown in Algorithms \ref{algo::ray-tracing-line-ratio-method1} (general ray-tracing steps) and \ref{algo::MIP0-line-ratio-method1} (MIP transfer function).

 \begin{algorithm}
  \caption{Line ratio (Method 1): outline of the front-to-back ray-tracing algorithm}
  \begin{algorithmic}[1]
    \Require{$cube1$: the first 3D array (strongest line); $cube2$: the second 3D array (faintest line); $step$: vector of length 3 representing ray increment; $loc$: vector of length 3 representing a location in 3D space; $N$: number of steps to take along the ray; $DAT\hspace{-0.1em}AMIN$: lower bound on data range.}
    \Statex
    \Function{rayTracing}{}
      \Let{$vars$}{initialise($vars$)}
      \Let{$loc$}{$startLoc$}
      \For{$i \gets 1 \textrm{ to } N$}
	\Let{$val1$}{$cube1[loc]$}
	\Let{$val2$}{$cube2[loc]$}
          \Let{$vars$}{transferFunction($vars$)}
          \Let{$loc$}{$loc$ + $step$}
      \EndFor
      \Let{$fragment$}{setFragment($vars$)}
      \State \Return{$fragment$}
    \EndFunction
  \end{algorithmic}
  \label{algo::ray-tracing-line-ratio-method1}
\end{algorithm}

\begin{algorithm}
  \caption{Line-Ratio (Method 1): MIP$_0$}
  \begin{algorithmic}[1]
    \Require{$val1$: value from first cube; $val2$: value from second cube; $maxval$: initialized to the smallest possible value; colourMap($scalar$): a function that maps a scalar to colour and transparency (RGBA).}
    \Statex
    \Function{transferFunction}{$val1, val2, maxval1, maxval2$}
	\If{$val1 > maxval1$}
		\Let{$maxval1$}{$val1$}
        \EndIf
	\If{$val2 > maxval2$}
		\Let{$maxval2$}{$val2$}
        \EndIf
      \State \Return{$maxval1, maxval2$}
    \EndFunction
    \\
    \Function{setFragment}{$maxval1, maxval2$}
    	\If{$maxval1>0.$} 
		\Let{$fragment$}{colourMap($\frac{maxval2}{maxval1}$)}
	\Else
		\Let{$fragment$}{colourMap($vec3(0., 0., 0., 0.)$)}
	\EndIf
      \State \Return{$fragment$}
    \EndFunction
  \end{algorithmic}
  \label{algo::MIP0-line-ratio-method1}
\end{algorithm}

{\bf Method 2.} As opposed to the previous method, 
the second method computes the ratio on a voxel by voxel basis. At each step taken by the ray, a ratio between the voxels from both cubes is computed, using this ratio as input for the transfer function. Since we compute the ratio on a voxel by voxel basis, this method has the potential to provide information about the spectral line shape (narrow vs broad), as the maximal ratio (with MIP) may be located away from the peaks of both lines. The method is shown in Algorithm \ref{algo::ray-tracing-line-ratio-method2} (general ray-tracing steps). For this scenario, MIP can be computed using Algorithm \ref{algo::MIP0} directly.

%
 \begin{algorithm}
  \caption{Line-Ratio (Method 2): outline of the front-to-back ray-tracing algorithm}
  \begin{algorithmic}[1]
    \Require{$cube1$: the first 3D array (strongest line); $cube2$: the second 3D array (faintest line); $step$: vector of length 3 representing ray increment; $loc$: vector of length 3 representing a location in 3D space; $N$: number of steps to take along the ray; $DAT\hspace{-0.1em}AMIN$: lower bound on data range.}
    \Statex
    \Function{rayTracing}{}
      \Let{$vars$}{initialise($vars$)}
      \Let{$loc$}{$startLoc$}
      \For{$i \gets 1 \textrm{ to } N$}
      	\If{$cube1[loc] > 0.$}
		\Let{$val$}{$\frac{cube2[loc]}{cube1[loc]}$}
	\Else
		\Let{$val$}{$DAT\hspace{-0.1em}AMIN$}
	\EndIf
          \Let{$vars$}{transferFunction($vars$)}
          \Let{$loc$}{$loc$ + $step$}
      \EndFor
      \Let{$fragment$}{setFragment($vars$)}
      \State \Return{$fragment$}
    \EndFunction
  \end{algorithmic}
  \label{algo::ray-tracing-line-ratio-method2}
\end{algorithm}

\section{Results and Discussion}
\label{sec::results_discussion}

In this section we demonstrate and compare the visual outcome of all transfer functions. Each transfer function is presented using both MIP and AVIP, along with filtering and smoothing. In addition, we demonstrate how graphic shaders can be used to explore datasets in real-time. 

\subsection{Software}
We implemented the transfer functions, filtering, and smoothing techniques in a custom standalone Python program, {\tt shwirl} \citep{Vohl2017ascl.soft04003V}. The program utilises Astropy \citep{Astropy2013A&A...558A..33A} to handle FITS files, Qt\footnote{\url{http://www.qtcentre.org}} (and PyQt) for the user interface, and VisPy \citep{campagnola:hal-01208191}, an object-oriented Python visualisation library binding onto OpenGL. We implemented the algorithms in the fragment shader using the GLSL language. 

{\tt Shwirl} has been developed primarily for the purpose of experimenting with shader algorithms. The software has been tested on Linux, Mac, and Windows machines, including remote desktop on cloud computing infrastructure. While the software is available for download and ready to visualise data, this is not intended as a full software release at the present time. 

\subsection{Test data}
\label{sec::testdata}
For our tests, we do not pre-process the data, but instead simply load it directly into GPU memory as a 3D texture. Doing so provides a way to highlight how shaders can be used to both process and visualise data at once. We use the following spectral cube data:

{\bf NGC 2903.} We select the THINGS spectral cube of NGC 2903 for it has the advantage of being well resolved in both spatial and spectral dimensions, enabling us to visualise noise and signal in the data\footnote{In particular, we use the beam corrected robust (ro) weighted data available at \url{http://www.mpia.de/THINGS/Data.html}.}. It is a good example of what is expected from upcoming radio survey data from APERTIF \citep[][]{Verheijen2009pra..confE..10V} and ASKAP \citep{Johnston2008ExA}. 

\begin{figure}
\centering
\includegraphics[width=8.3cm]{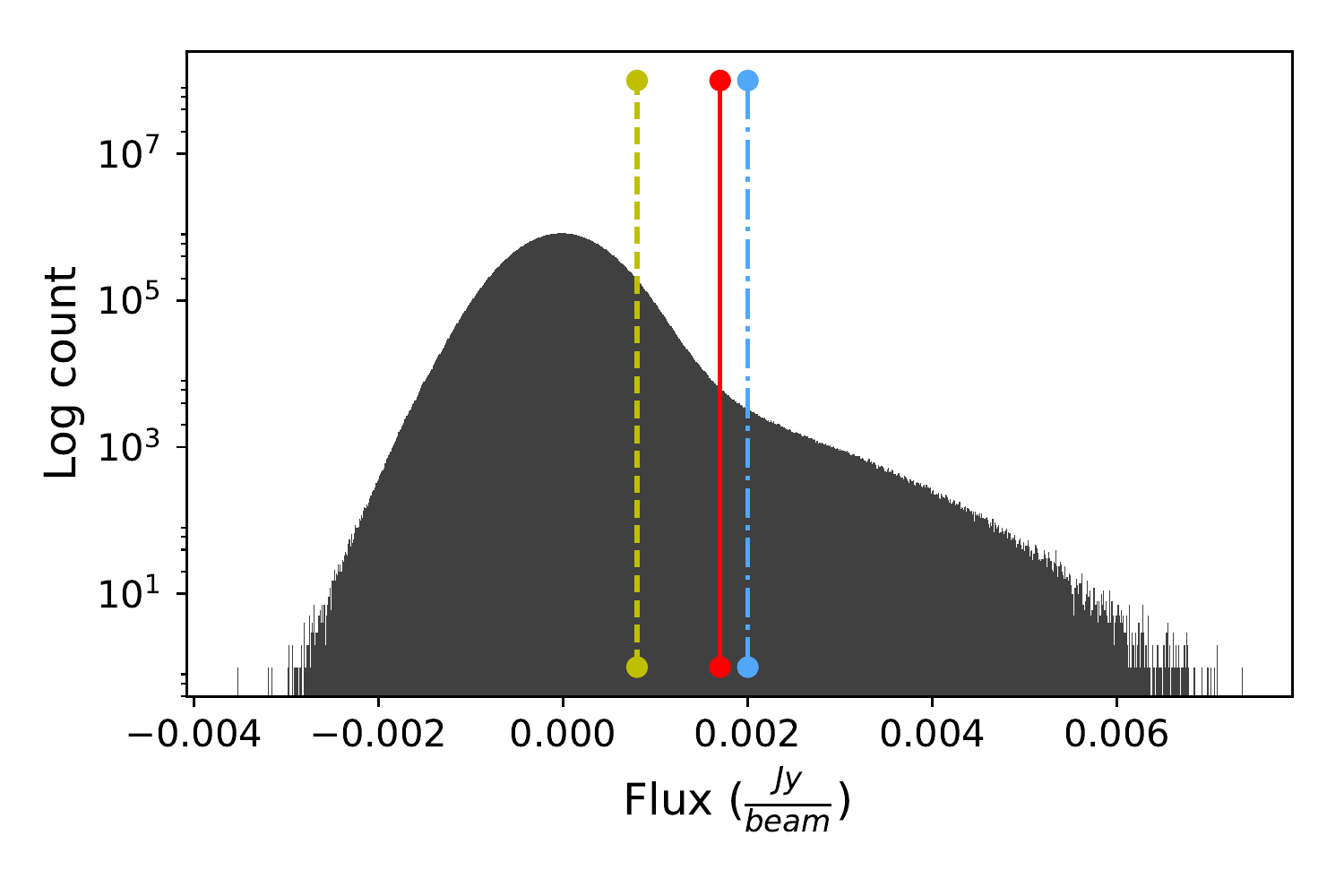}
\caption{Histogram of voxel flux intensity from the NGC 2903 cube \citep{Walter2008AJ}. Data minimum=$-3.5\times10^{-3}$ Jy/beam; maximum=$7.4\times10^{-3}$ Jy/beam; and standard deviation=$0.5\times10^{-3}$ Jy/beam. The vertical lines show the minimum threshold parameters used in Figures \ref{fig::compare1} and \ref{fig::compare2} (red, solid: $minT\hspace{-0.1em}hreshold=1.7\times10^{-3}$ Jy/beam, yellow, dashed: $minT\hspace{-0.1em}hreshold=0.8\times10^{-3}$ Jy/beam, and blue, dotted-dashed: $minT\hspace{-0.1em}hreshold=2\times10^{-3}$Jy/beam). Cube dimensions (ra, dec, vel) in voxels: (1024, 1024, 87).}

\label{fig::histogram-things}
\end{figure}

Velocity channels are Doppler-shifted to the barycentric frame (using the FELO-HEL convention originating from AIPS\footnote{\url{http://www.aips.nrao.edu/index.shtml}} --- a regular grid in frequency) and expressed in unit of kilometres per second (km/s). Figure \ref{fig::histogram-things} shows the histogram of voxel intensity in the NGC 2903 cube.

{\bf Antennae galaxies.} This spectral cube is the Southern mosaic pattern taken from the ALMA Science Verification data targeting the CO 3-2 line in the Antennae galaxies\footnote{See \url{https://almascience.nao.ac.jp/alma-data/science-verification/Antennae-galaxies} for more details.}. We select this dataset for it is well resolved both spatially and spectrally, and is a representative data product of current and upcoming CO-related studies. In addition, it enables us to evaluate how the transfer functions and shaders behave with mosaic data (e.g. a sparse cube where some voxels do not contain information about the observation). 

\begin{figure}
\centering
\includegraphics[width=8.3cm]{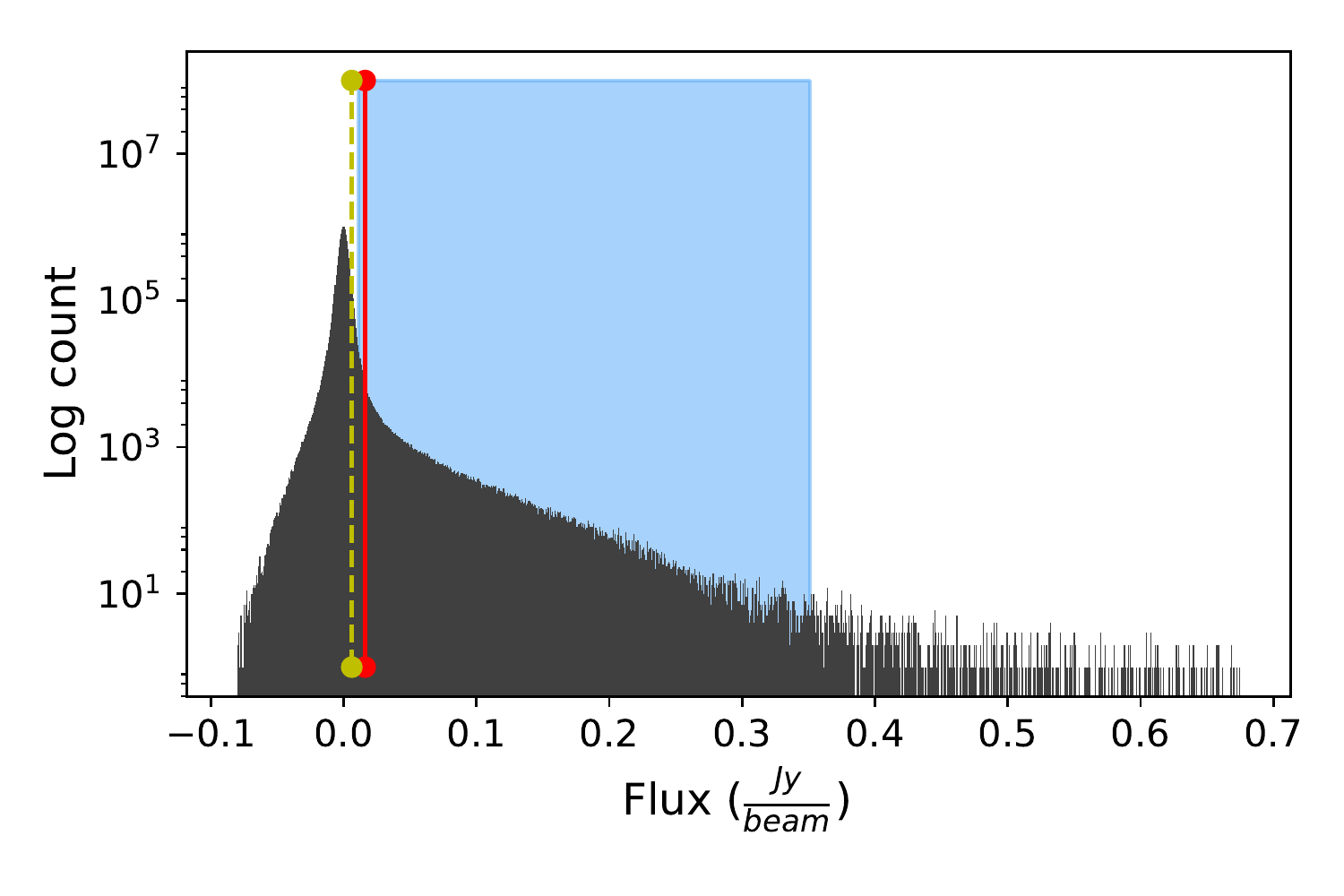}
\caption{Histogram of voxel flux intensity from the Antennae cube. Data minimum=$-8\times10^{-2}$ Jy/beam; maximum=$6.8\times10^{-1}$ Jy/beam; and standard deviation=$1\times10^{-2}$. NAN values (distributed around the tilted polygon --- see Figure \ref{fig::compare3}) are discarded from the histogram. The vertical lines show the minimum threshold parameters used in Figures \ref{fig::compare3} and \ref{fig::compare4} (red, solid: $minT\hspace{-0.1em}hreshold=0.63\times10^{-2}$ Jy/beam, yellow, dashed: $minT\hspace{-0.1em}hreshold=1.63\times10^{-2}$ Jy/beam). The blue region highlights the regions between $minT\hspace{-0.1em}hreshold$ and $maxT\hspace{-0.1em}hreshold$ used for intensity domain scaling. Cube dimensions (ra, dec, vel) in voxels: (750, 750, 70).}

\label{fig::histogram-Antennae}
\end{figure}

The third dimension represents frequency expressed in units of tera-Hertz (THz). Figure \ref{fig::histogram-Antennae} shows the histogram of voxel intensity in the Antennae cube. 

{\bf GAMA-511867.} The two sub cubes of the barred galaxy GAMA-511867 have been extracted from the ``red'' spectral cube\footnote{Available from \url{https://sami-survey.org/edr/browser}} from the SAMI survey, obtained with the Sydney-AAO Multi-object Integral field spectrograph \citep{Croom2012MNRAS}. The two cubes correspond to the H$\alpha$ and [NII] emission lines. The cubes have been selected as both emission lines have a high signal to noise.

\begin{figure}
\centering
\includegraphics[width=8.3cm]{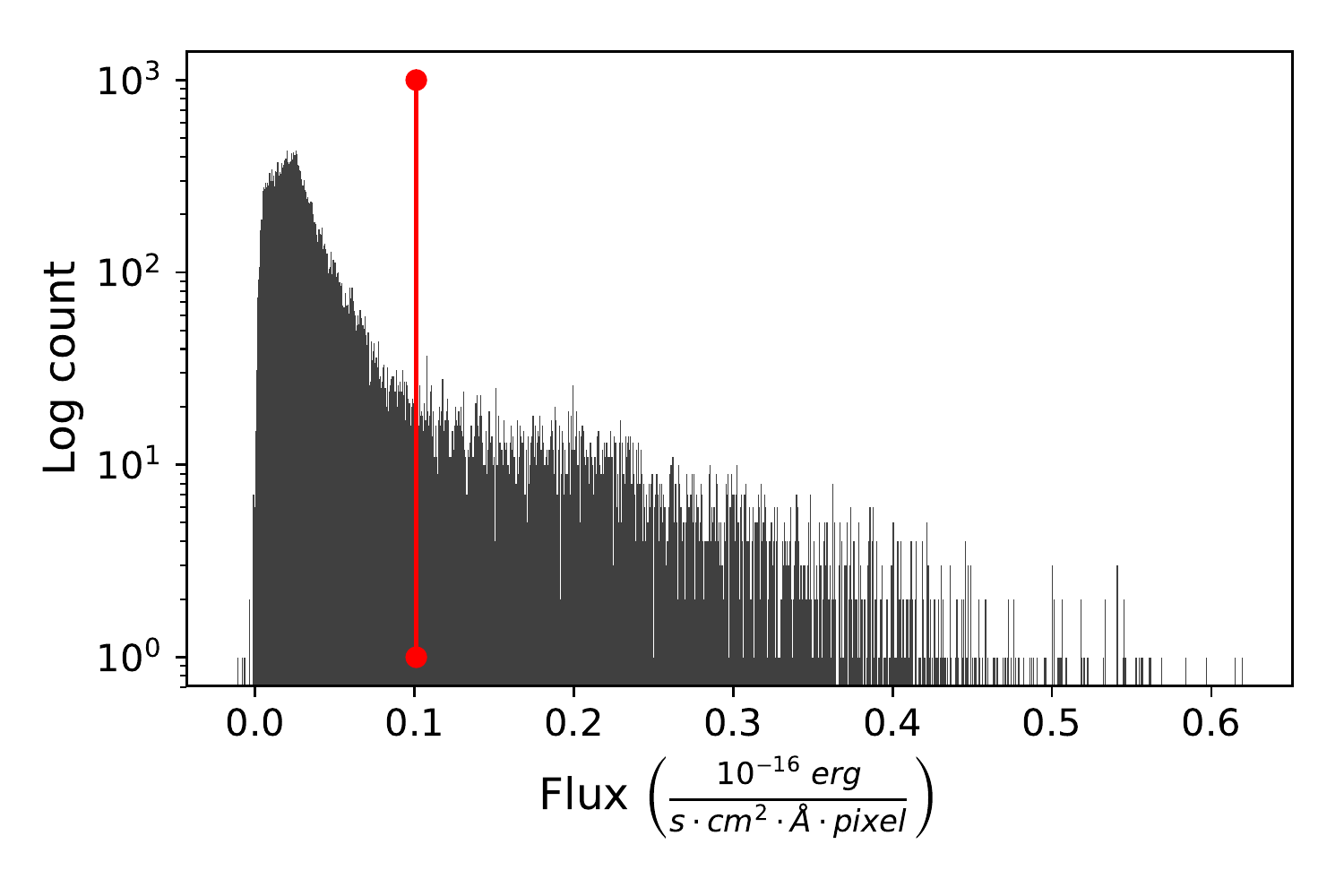}
\caption{Histogram of voxel flux intensity from the H$\alpha$ cube of GAMA-511867. Data minimum=$-1.14\times10^{-2} \times 10^{-16}$ erg s$^{-1}$ cm$^{-2}$ \r{A}$^{-1}$ pixel$^{-1}$; maximum=$6.2\times10^{-1} \times 10^{-16}$ erg s$^{-1}$ cm$^{-2}$ \r{A}$^{-1}$ pixel$^{-1}$; and standard deviation=$7.68\times10^{-2}$. NAN values (distributed around the tilted polygon --- see Figure \ref{fig::compare3}) are discarded from the histogram. The vertical red line shows the minimum threshold parameter used in Figure \ref{fig::line-ratio2} ($minT\hspace{-0.1em}hreshold=0.1011 \times 10^{-16}$ erg s$^{-1}$ cm$^{-2}$ \r{A}$^{-1}$ pixel$^{-1}$). Cube dimensions (ra, dec, $\lambda$) in voxels correspond to (50, 50, 37).}

\label{fig::histogram-sami-halpha}
\end{figure}

\begin{figure}
\centering
\includegraphics[width=8.3cm]{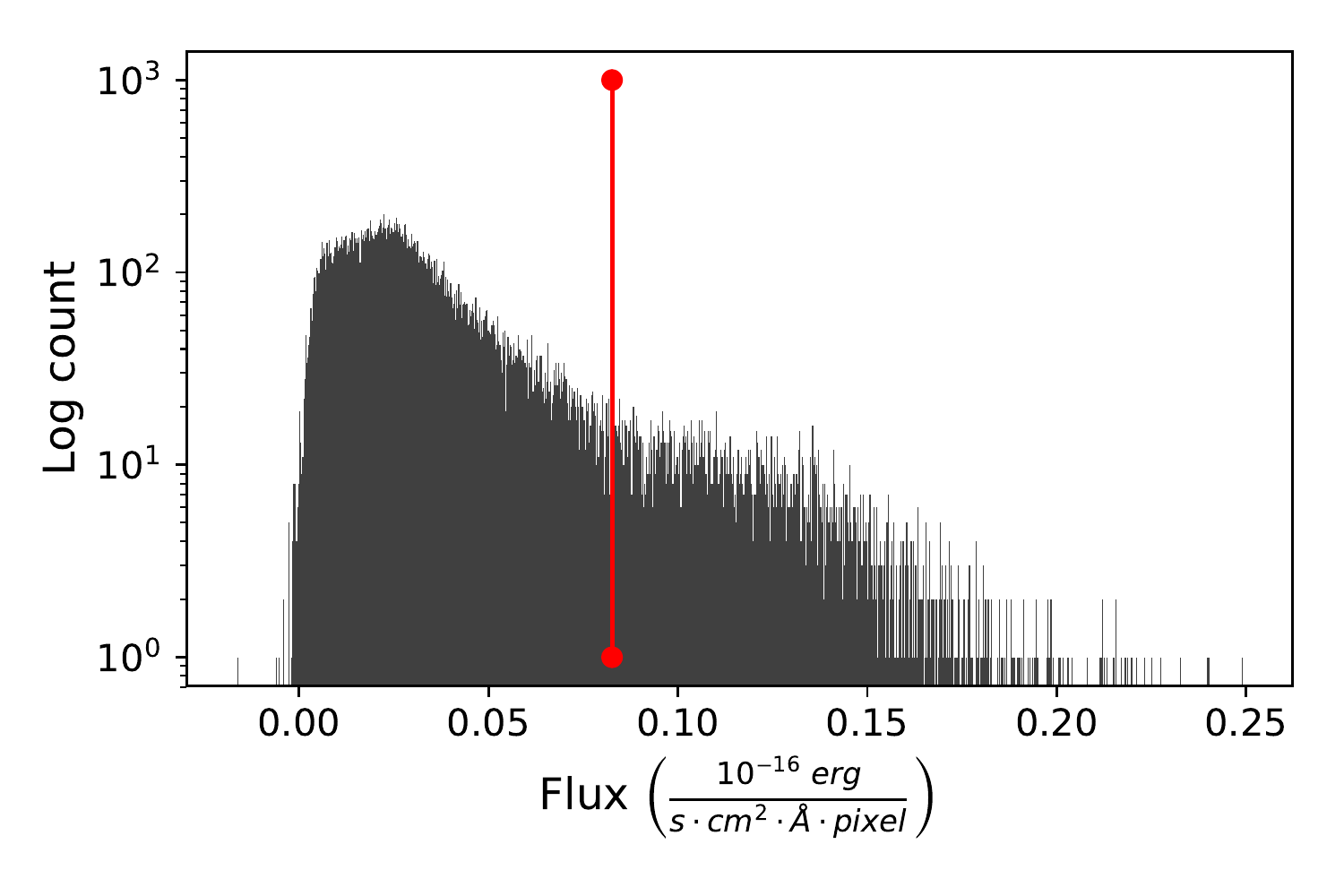}
\caption{Histogram of voxel flux intensity from the H$\alpha$ cube of GAMA-511867. Data minimum=$-1.64\times10^{-2} \times 10^{-16}$ erg s$^{-1}$ cm$^{-2}$ \r{A}$^{-1}$ pixel$^{-1}$; maximum=$2.49\times10^{-1} \times 10^{-16}$ erg s$^{-1}$ cm$^{-2}$ \r{A}$^{-1}$ pixel$^{-1}$; and standard deviation=$3.29\times10^{-2}$. NAN values (distributed around the tilted polygon --- see Figure \ref{fig::compare3}) are discarded from the histogram. The vertical red line shows the minimum threshold parameter used in Figure \ref{fig::line-ratio2} ($minT\hspace{-0.1em}hreshold=0.0826 \times 10^{-16}$ erg s$^{-1}$ cm$^{-2}$ \r{A}$^{-1}$ pixel$^{-1}$). Cube dimensions (ra, dec, $\lambda$) in voxels correspond to (50, 50, 37).}

\label{fig::histogram-sami-nii}
\end{figure}

The third dimension represents wavelength, expressed in units of \r{A}ngstr\"{o}m (\r{A}). Figures \ref{fig::histogram-sami-halpha} and \ref{fig::histogram-sami-nii} show the histogram of voxel flux intensity in the H$\alpha$ and [NII] cubes respectively. 

\subsection{Results}
In this section, we present a qualitative evaluation of the transfer functions. The result of each transfer function (Algorithms \ref{algo::MIP0} to \ref{algo::RGB-AVIP}), with and without filtering and smoothing, is presented in Figures \ref{fig::compare1}, \ref{fig::compare2} (NGC 2903 cube), \ref{fig::compare3}, and \ref{fig::compare4} (Antennae cube). We hide axis labels to emphasise the different type of information provided by the colouring methods. 

Figures \ref{fig::compare1} and \ref{fig::compare3} show volume rendering using parallel projection, with spectral cube view face-on (i.e. orthogonal to the spectral axis). Figures \ref{fig::compare2} and \ref{fig::compare4} show volume rendering using perspective projection, with the camera position set 
to provide a blend of spatial and spectral information. Parallel projection figures are presented to highlight the resemblance to the outcome of the ``moment-inspired'' transfer functions with the zeroth and first moment maps. Perpective projection figures are presented to provide an example of the type of visualisation suitable for stereoscopic display. 

In Figures \ref{fig::compare1} to \ref{fig::compare4}, each column displays a specific transfer function: left, center, and right columns show the zeroth moment-inspired, first moment-inspired, and RGB transfer functions respectively. At the top of the figures the colour maps are shown for the individual panels (left and right columns)\footnote{Note that when using AVIP, the label should be Jy beam$^{-1}$ km s$^{-1}$ and should consider the weighting factor $k$ (see, for example, algorithm \ref{algo::AVIP0}). For the simplicity of the figure, we only show Jy/beam.}, and the RGB cube is also provided to help with the interpretation of colours location in the 3D RGB space. As discussed in Section \ref{sec::rgb}, the RGB transfer function provides a visual cue of the 3D distribution of emission, as each colour corresponds to a specific location in the 3D RGB space. Each row shows a different combination of transfer function with different visualisation settings (MIP, AVIP, filtering and smoothing).

\subsection{The NGC 2903 cube}

\begin{figure*}
\centering
\includegraphics[width=13cm]{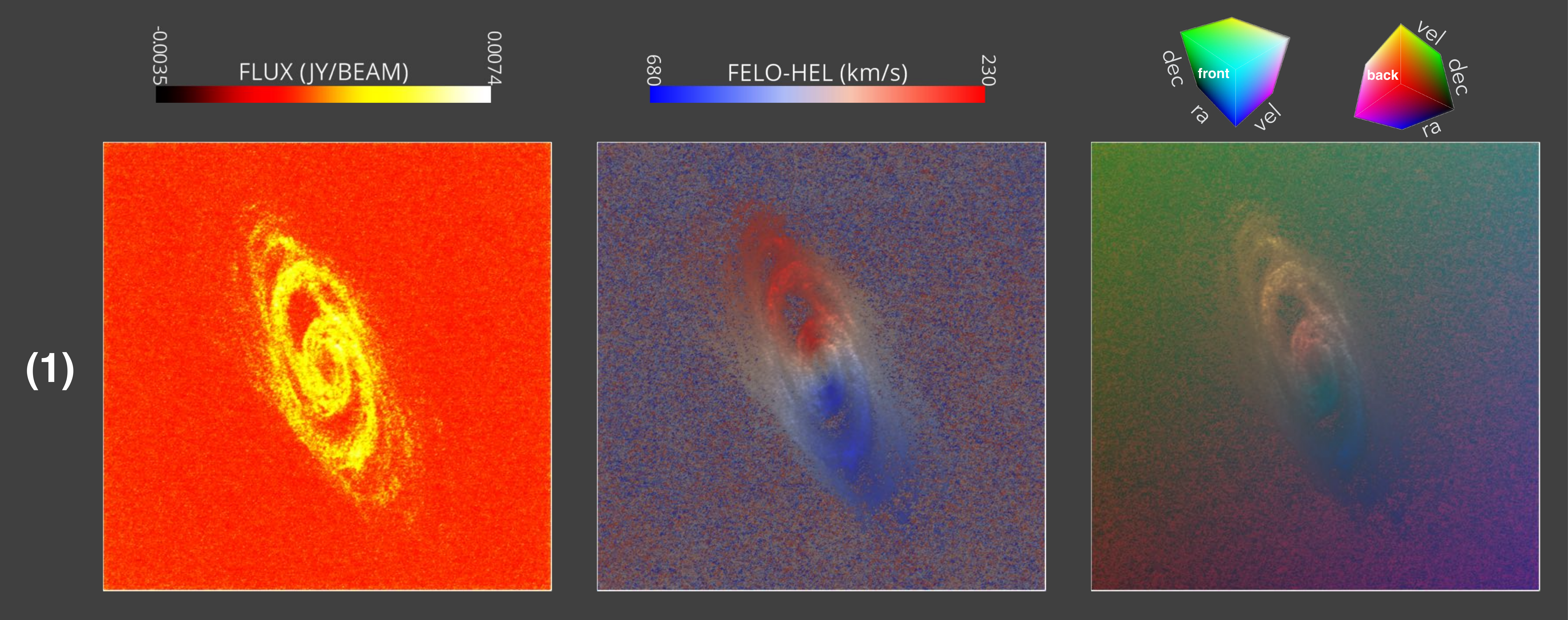}\\
\vspace{-0.045cm}
\includegraphics[width=13cm]{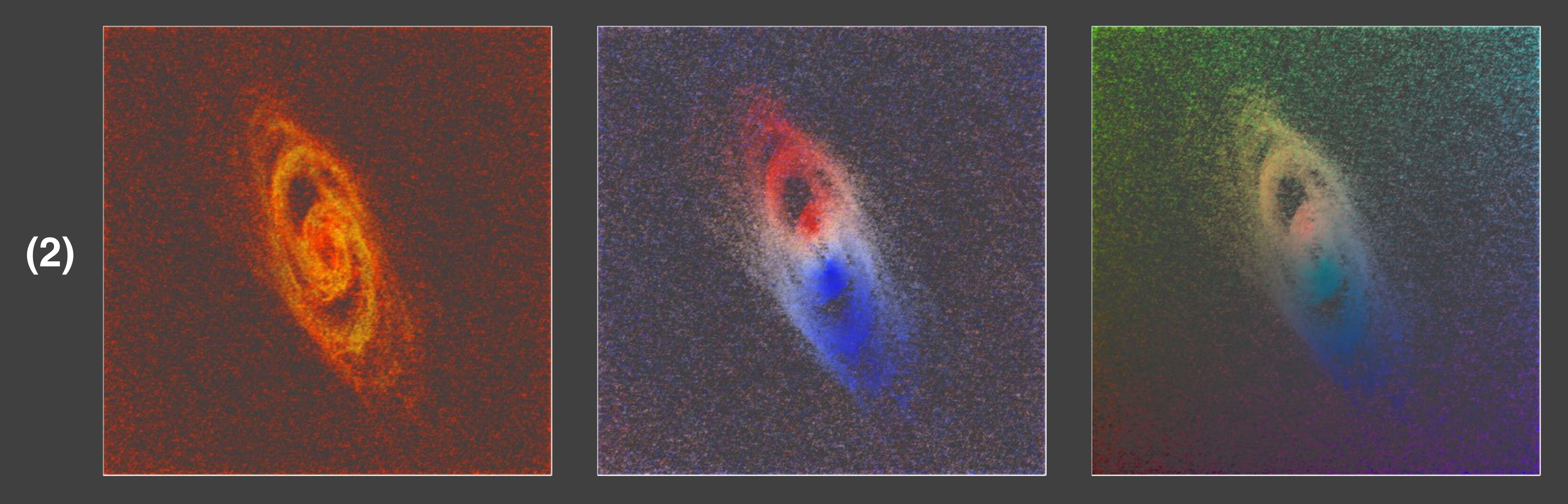}\\
\vspace{-0.05cm}
\includegraphics[width=13cm]{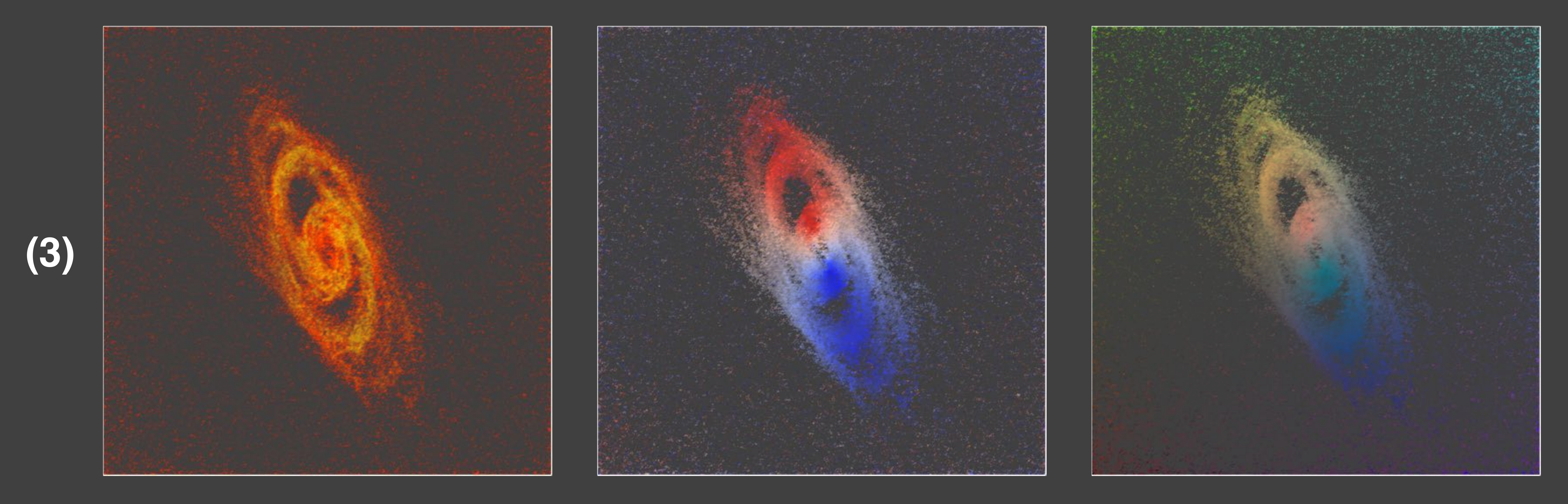}\\
\vspace{-0.05cm}
\includegraphics[width=13cm]{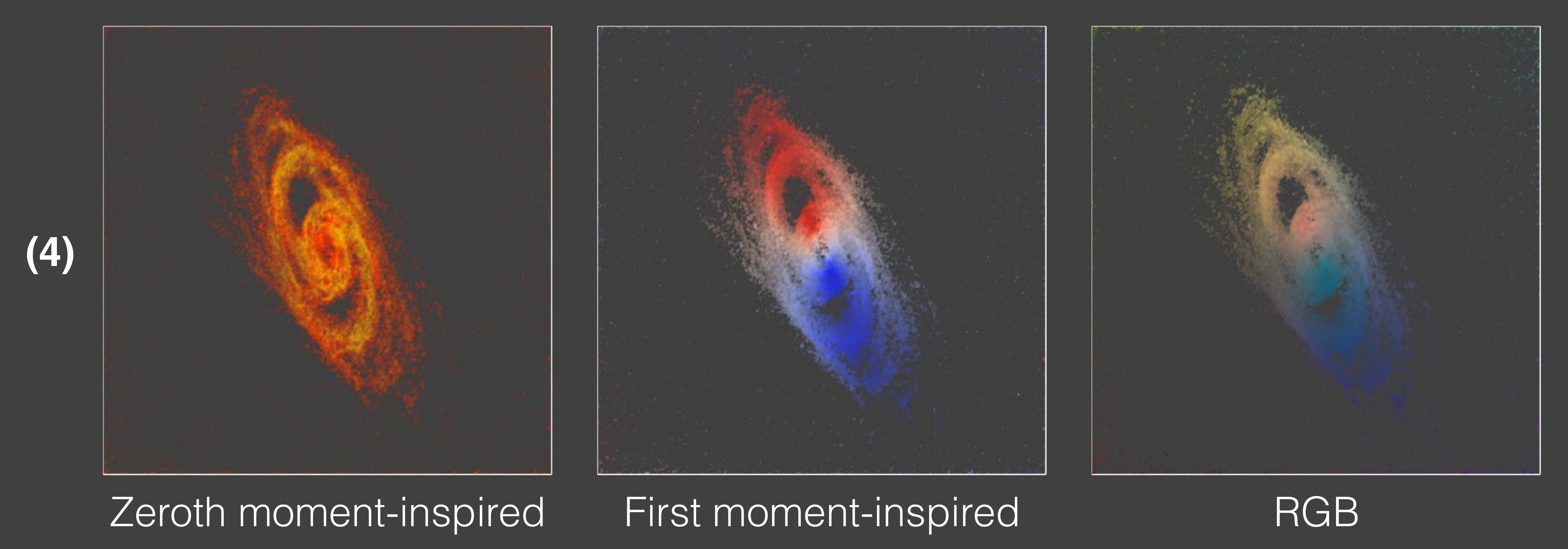}\\
\caption{Comparison between transfer functions using parallel projection for the NGC 2903 cube. (1) MIP without filtering or smoothing; (2) AVIP with intensity clipping ($minT\hspace{-0.1em}hreshold=0.8\times10^{-3}$ Jy/beam); (3) AVIP with intensity clipping and box smoothing ($filterArm=1$); and (4) AVIP with intensity clipping and 9-tap Gaussian smoothing. For all AVIP, $k=0.31$ (weighting factor).}

\label{fig::compare1}
\end{figure*}

Figure \ref{fig::compare1} displays the NGC 2903 spectral cube viewed face-on, with parallel projection. The four rows show combinations of different transfer functions : 
\renewcommand{\theenumi}{(\arabic{enumi})}
\begin{enumerate}
\item MIP without any filtering or smoothing --- therefore purely rendering the spectral cube;
\item AVIP using a weighting factor of $k=0.31$, with intensity clipping ($minT\hspace{-0.1em}hreshold=0.8\times10^{-3}$ Jy/beam); 
\item AVIP with intensity clipping [same $k$ and $minT\hspace{-0.1em}hreshold$ as (2)] and box smoothing (a box of size 3 --- $filterArm=1$);
\item AVIP with intensity clipping [same $k$ and $minT\hspace{-0.1em}hreshold$ as (2)] and 9-tap Gaussian smoothing.
\end{enumerate}

\begin{figure*}
\centering
\includegraphics[width=13cm]{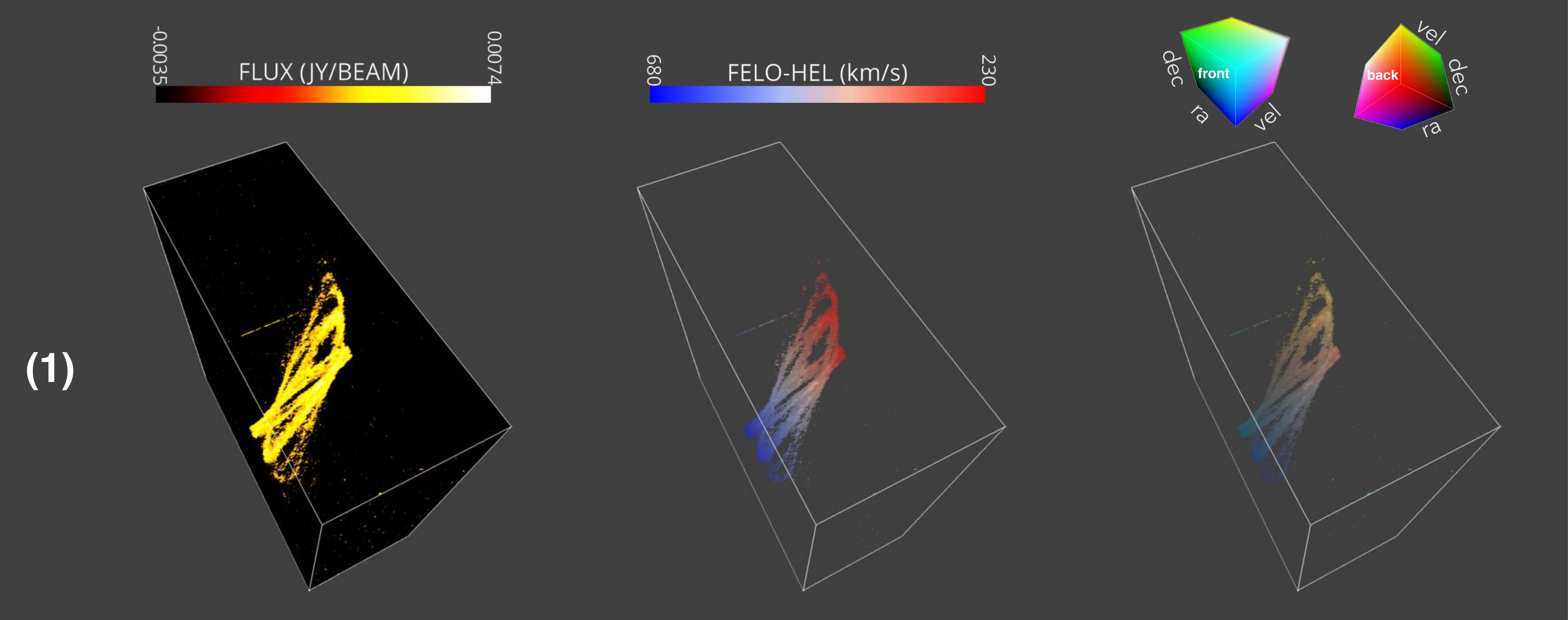}\\
\vspace{-0.045cm}
\includegraphics[width=13cm]{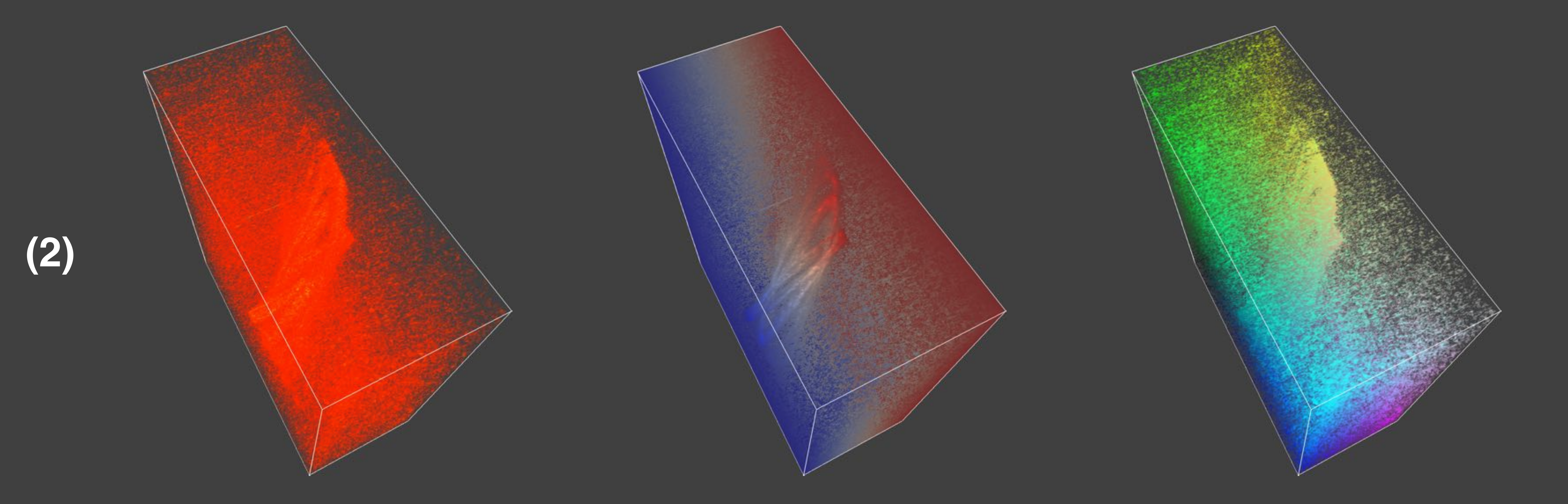}\\
\vspace{-0.05cm}
\includegraphics[width=13cm]{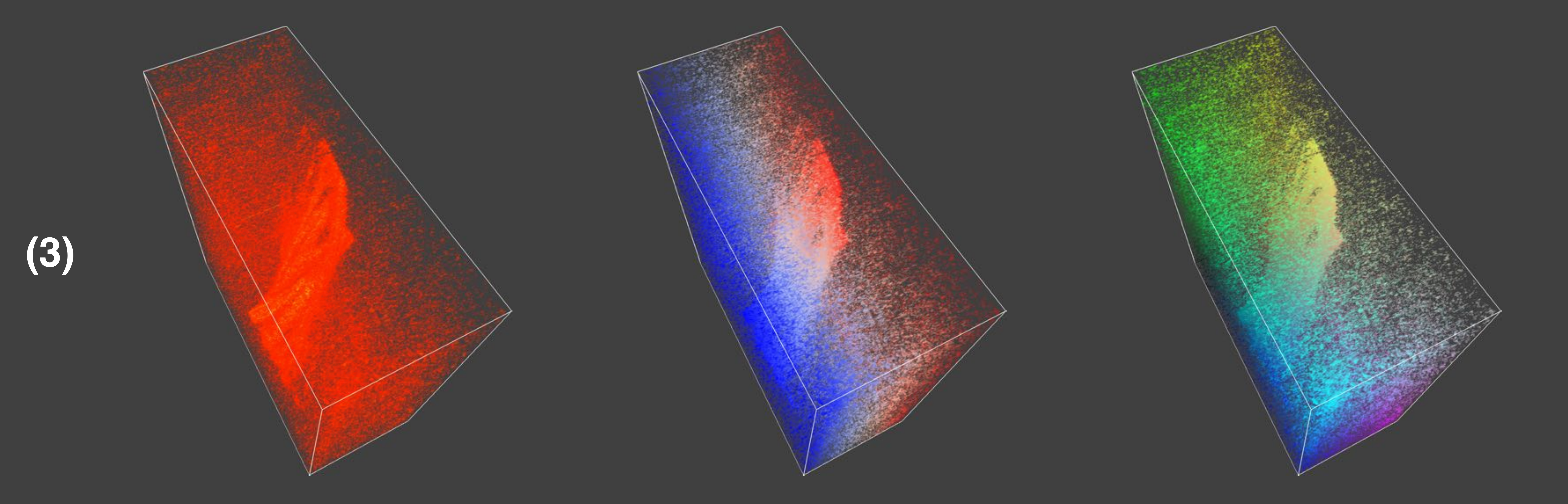}\\
\vspace{-0.05cm}
\includegraphics[width=13cm]{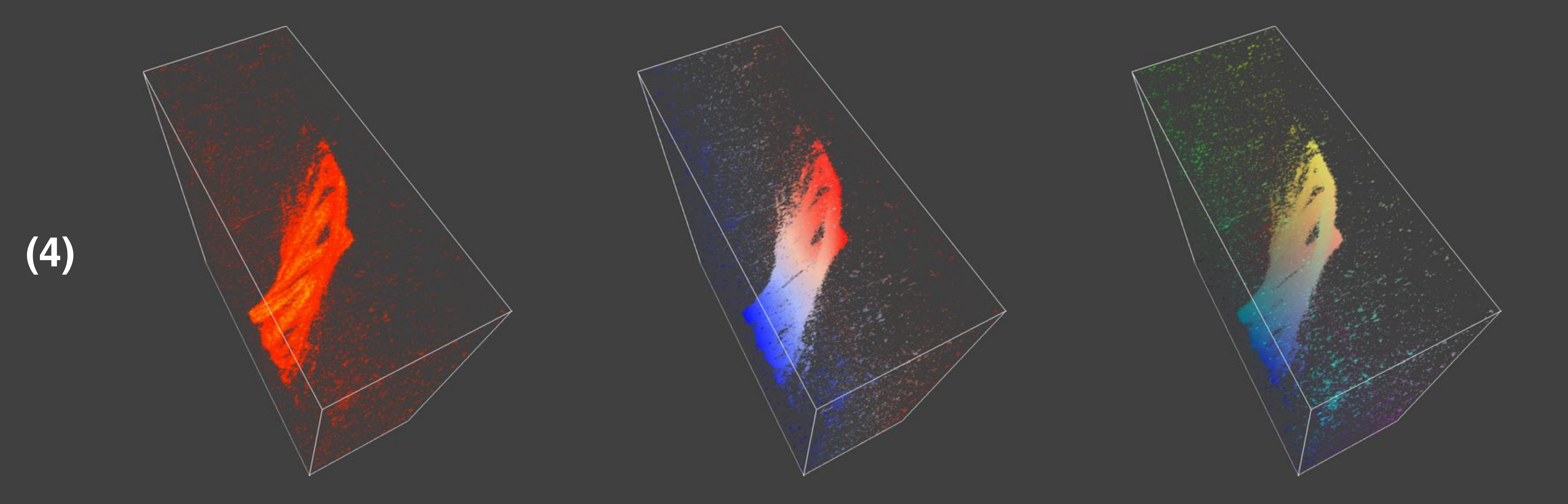}\\
\vspace{-0.05cm}
\includegraphics[width=13cm]{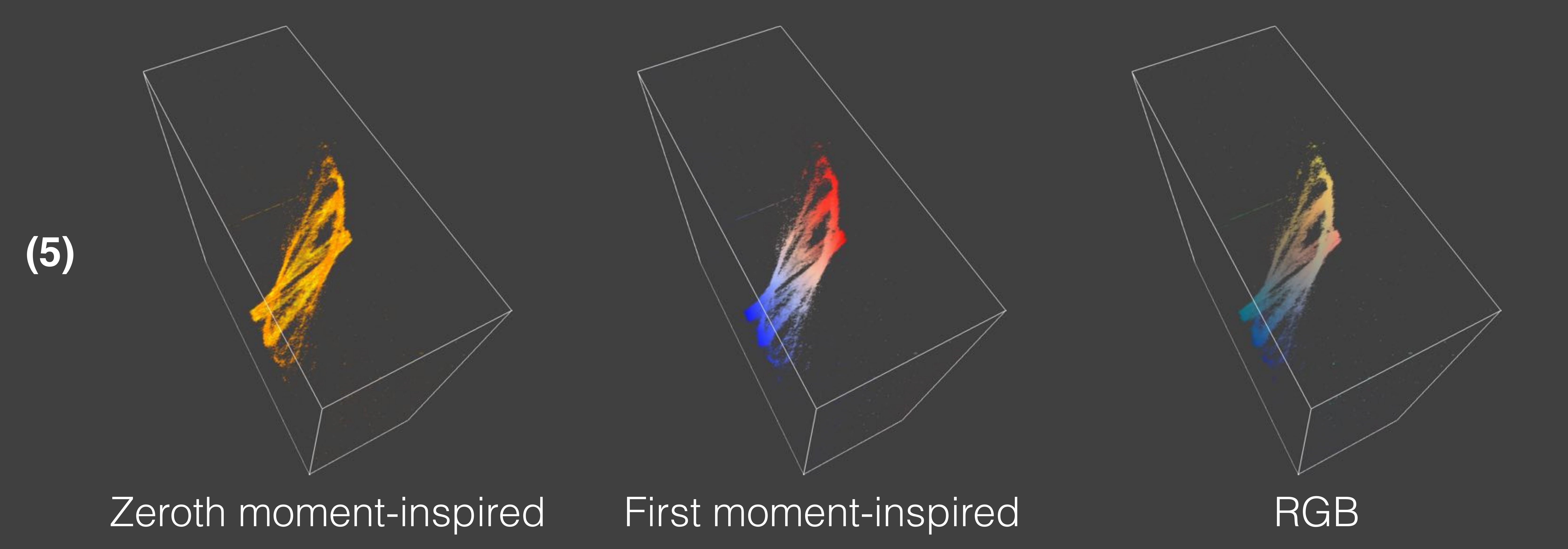}\\
\caption{Comparison between transfer functions using perspective projection for the NGC 2903 cube. (1) MIP with intensity clipping ($minT\hspace{-0.1em}hreshold=2\times10^{-3}$ Jy/beam); (2) AVIP with intensity clipping ($minT\hspace{-0.1em}hreshold=0.8\times10^{-3}$ Jy/beam); (3) AVIP with intensity clipping and box smoothing ($filterArm=1$); (4) AVIP with intensity clipping and 9-tap Gaussian smoothing; and (5) AVIP with $k=0.22$, intensity clipping ($minT\hspace{-0.1em}hreshold=1.7\times10^{-3}$ Jy/beam), and 5-tap Gaussian smoothing. For AVIP in (2), (3) and (4), $k=0.31$.}
\label{fig::compare2}
\end{figure*}

Figure \ref{fig::compare2} presents the same set of parameters as in Figure \ref{fig::compare1}, with the exception of the fifth row, which shows the combination of transfer functions with AVIP ($k=0.22$), intensity clipping ($minT\hspace{-0.1em}hreshold=1.7\times10^{-3}$ Jy/beam), and 5-tap Gaussian smoothing. The fifth row is added to provide a sense of what can be achieved through different sets of visualisation parameters.  

All visualisation parameters (e.g. $minT\hspace{-0.1em}hreshold$, $k$, $filterArm$) can be selected and modified interactively at run-time, in order to let the user explore and highlight features of interest in the data. Therefore, for the figures, we selected parameters that highlight properties of each technique. For instance, we set the value of $k$ (for rows 2 to 4) in order to keep some of the low intensity noise around the source. This further allows us to show the effect of smoothing in rows 3 and 4. 

\subsection{The Antennae cube}

\begin{figure*}
\centering
\includegraphics[width=13cm]{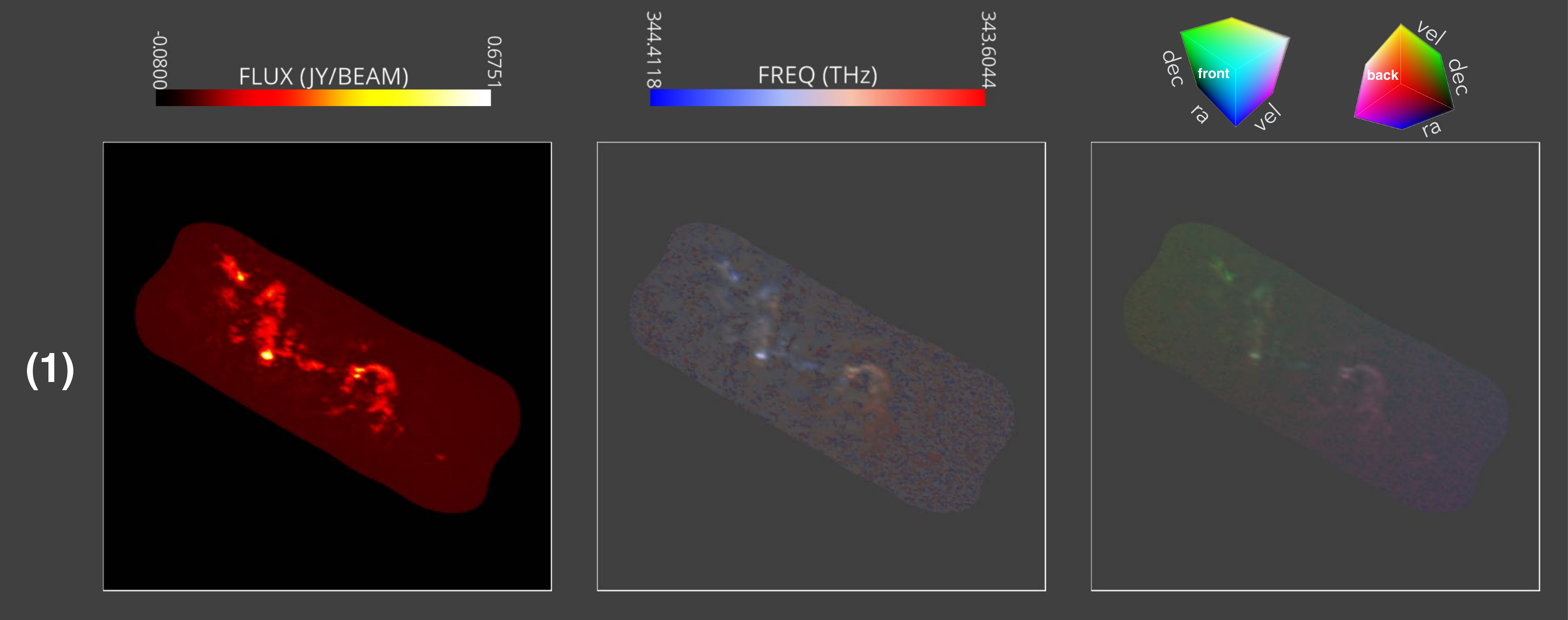}\\
\vspace{-0.045cm}
\includegraphics[width=13cm]{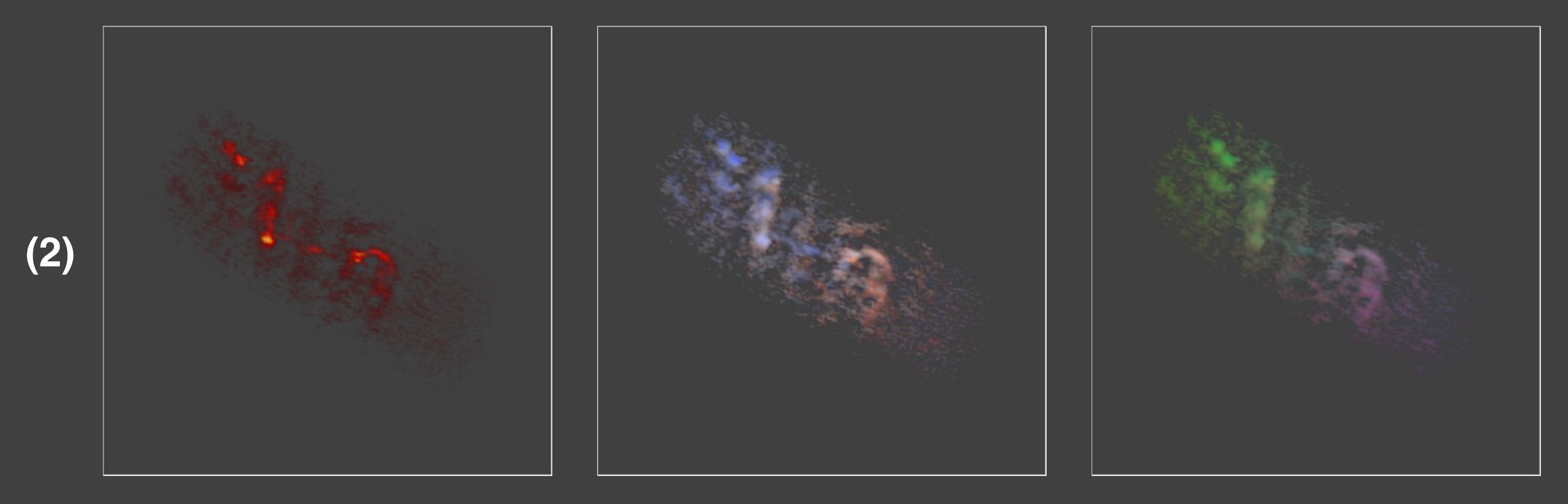}\\
\vspace{-0.05cm}
\includegraphics[width=13cm]{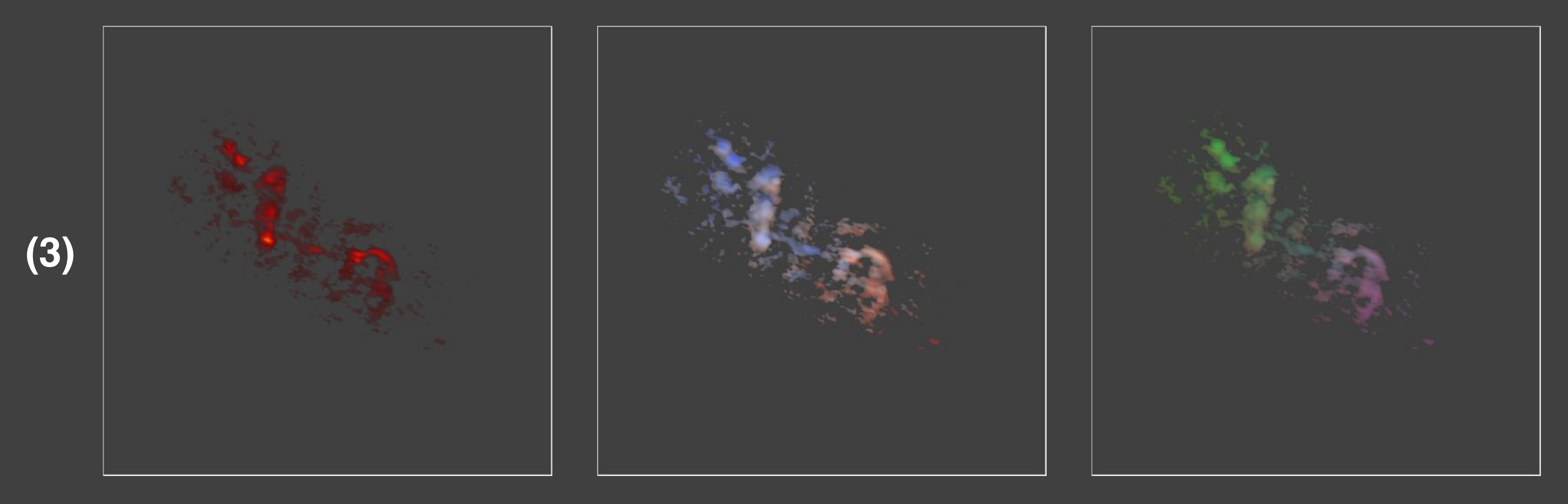}\\
\vspace{-0.05cm}
\includegraphics[width=13cm]{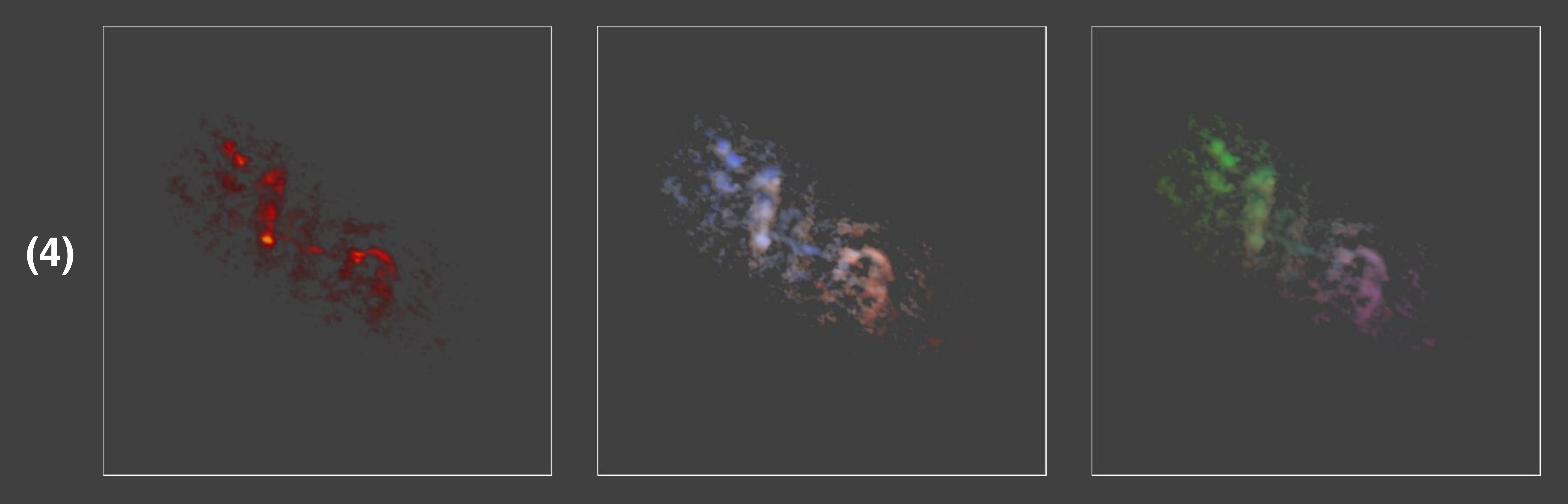}\\
\vspace{-0.05cm}
\includegraphics[width=13cm]{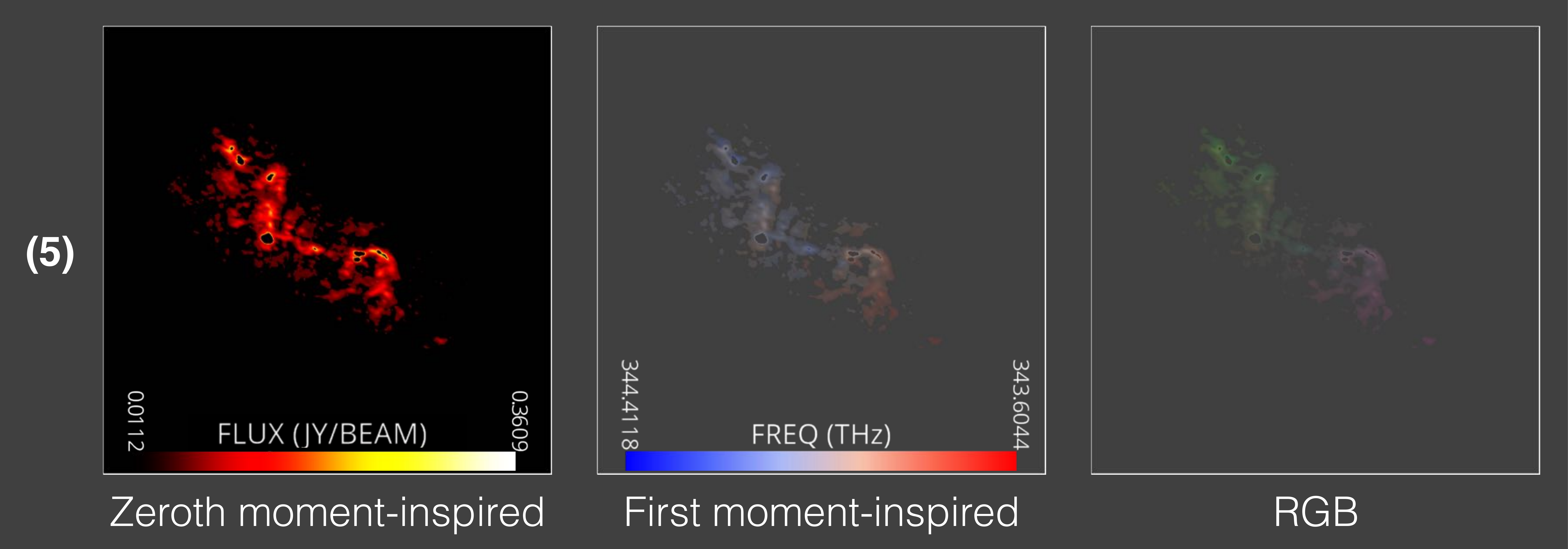}\\
\caption{Comparison between transfer functions using parallel projection for the Antennae cube. (1) MIP without filtering nor smoothing; (2) AVIP with intensity clipping ($minT\hspace{-0.1em}hreshold=0.62\times10^{-2}$ Jy/beam); (3) AVIP with intensity clipping and box smoothing ($filterArm=5$); (4) AVIP with intensity clipping and 9-tap Gaussian smoothing; and (5) MIP with intensity domain scaling (Algorithm \ref{algo::scale};  $minT\hspace{-0.1em}hreshold=1.12\times10^{-2}$ Jy/beam, $maxT\hspace{-0.1em}hreshold=0.3609$ Jy/beam), and a 9-tap Gaussian smoothing. For all AVIP, $k=0.31$ (weighting factor).}
\label{fig::compare3}
\end{figure*}

Figure \ref{fig::compare3} displays the Antennae spectral cube viewed face-on, with parallel projection, and where the four rows show the transfer function combined to : 
\begin{enumerate}
\item MIP without any filtering or smoothing; 
\item AVIP using a weighting factor of $k=0.31$, and intensity clipping ($minT\hspace{-0.1em}hreshold=0.62\times10^{-2}$ Jy/beam); 
\item AVIP with intensity clipping [same $k$ and $minT\hspace{-0.1em}hreshold$ as (2)] and box smoothing (a box of size 9); 
\item AVIP with intensity clipping [same values of $k$ and $minT\hspace{-0.1em}hreshold$ as in (2)] and 9-tap Gaussian smoothing. 
\end{enumerate}

\begin{figure*}
\centering
\includegraphics[width=13cm]{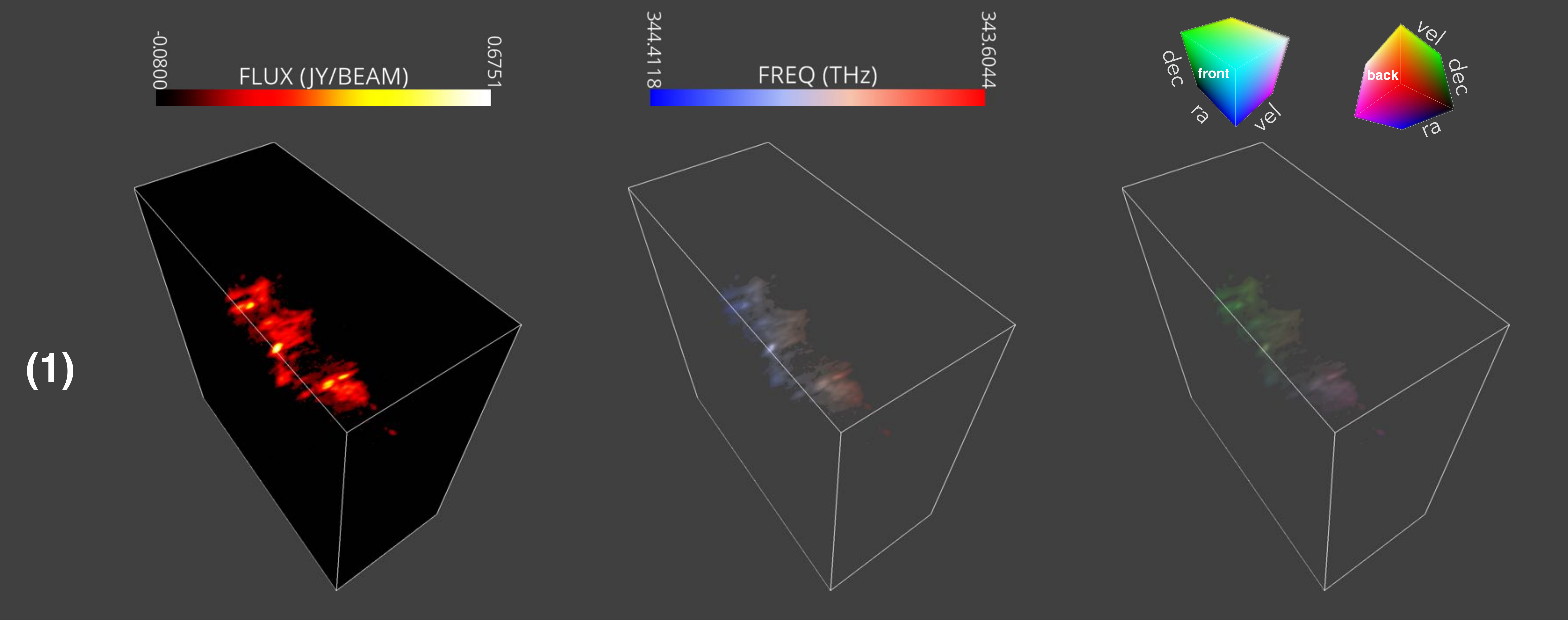}\\
\vspace{-0.045cm}
\includegraphics[width=13cm]{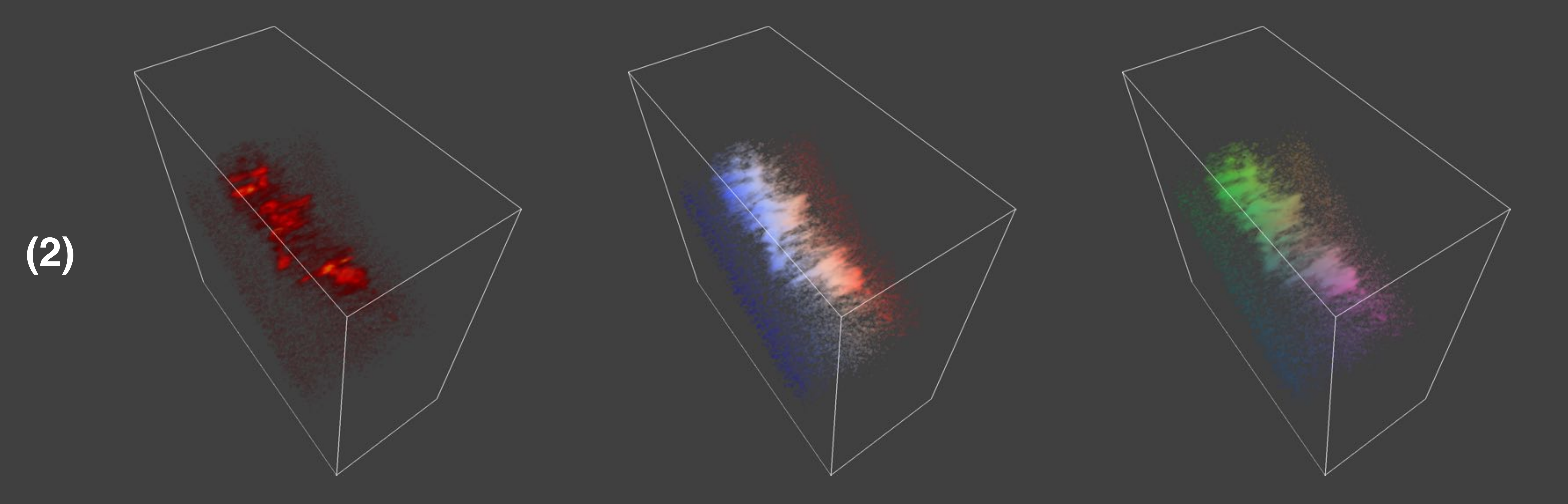}\\
\vspace{-0.05cm}
\includegraphics[width=13cm]{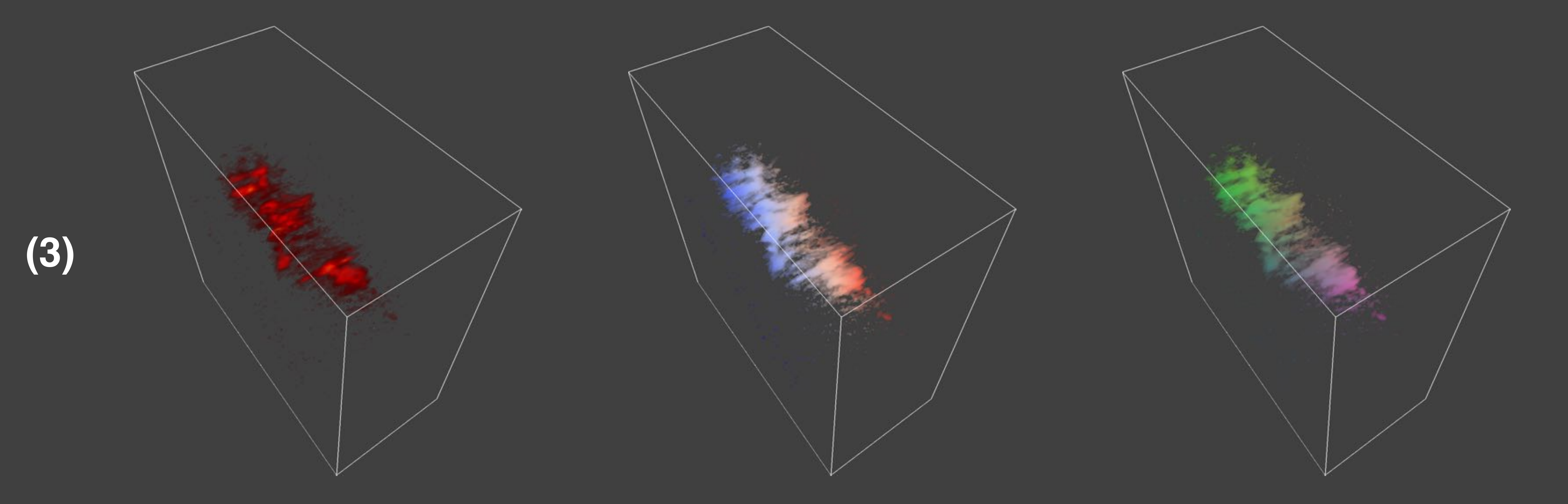}\\
\vspace{-0.05cm}
\includegraphics[width=13cm]{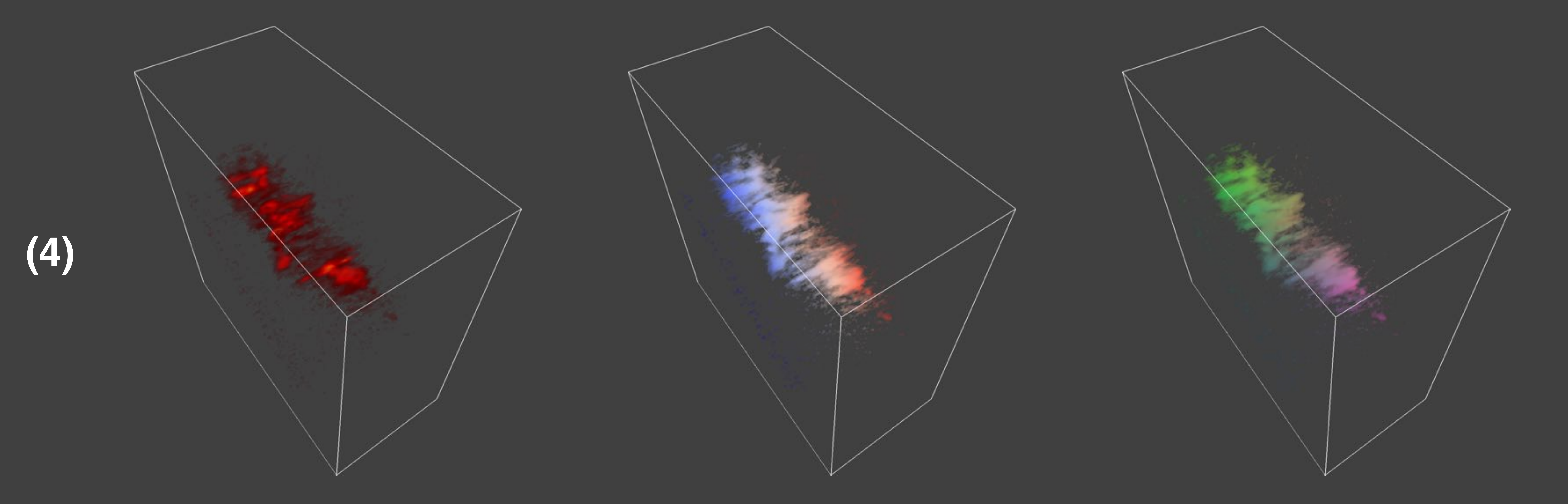}\\
\vspace{-0.05cm}
\includegraphics[width=13cm]{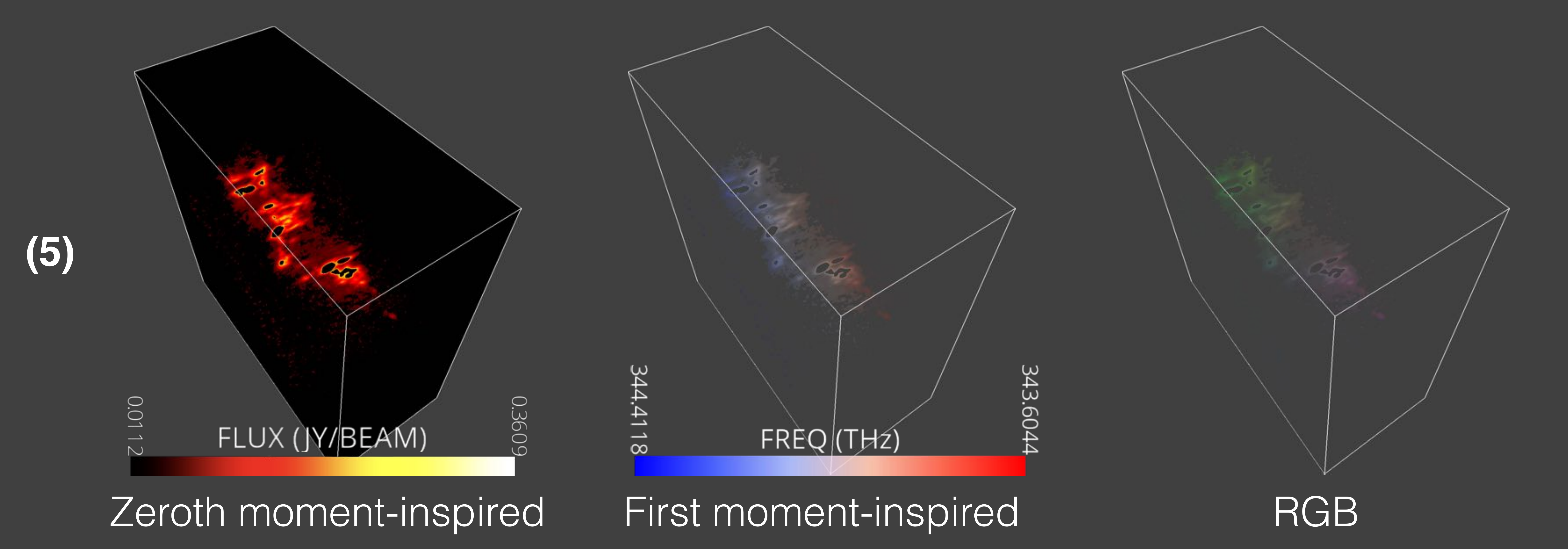}\\
\caption{Comparison between transfer functions using perspective projection for the Antennae cube. (1) MIP with intensity clipping ($1.63\times10^{-2}$) and 9-tap Gaussian smoothing; (2) AVIP with intensity clipping ($minT\hspace{-0.1em}hreshold=0.62\times10^{-2}$ Jy/beam); (3) AVIP with intensity clipping and box smoothing ($filterArm=4$); (4) AVIP with intensity clipping and 9-tap Gaussian smoothing; and (5) MIP with intensity domain scaling (Algorithm \ref{algo::scale};  $minT\hspace{-0.1em}hreshold=1.12\times10^{-2}$ Jy/beam, $maxT\hspace{-0.1em}hreshold=0.3609$ Jy/beam), and a 9-tap Gaussian smoothing. For AVIP in (2), (3), and (4), $k$ is set to the same value as in Figure \ref{fig::compare3}.}
\label{fig::compare4}
\end{figure*}

Figure \ref{fig::compare4} presents the same set of parameters as in Figure \ref{fig::compare3}, with the exception of the first and last rows. In the first row, we now use MIP with intensity clipping ($minT\hspace{-0.1em}hreshold=1.63\times10^{-2}$ Jy/beam) and 9-tap Gaussian smoothing. The threshold highlights only the strongest emission, and removes the polygon emerging from the mosaic. 

In the last row, we show MIP with intensity domain scaling (Algorithm \ref{algo::scale}), where:
\begin{itemize}
\item $minT\hspace{-0.1em}hreshold=1.12\times10^{-2}$ Jy/beam,
\item $maxT\hspace{-0.1em}hreshold=0.3609$ Jy/beam; 
\item and 9-tap Gaussian smoothing.
\end{itemize} 

The intensity domain scaling shows local maxima in the data by discarding the very bright regions, which are being masked in the visualisation. Note that the colour map for the first moment-inspired transfer function did not change while the range for the zeroth moment-inspired did change [provided along row (5)].

\subsection{The GAMA-511867 cubes}

\begin{figure*}
\centering
\includegraphics[width=13cm]{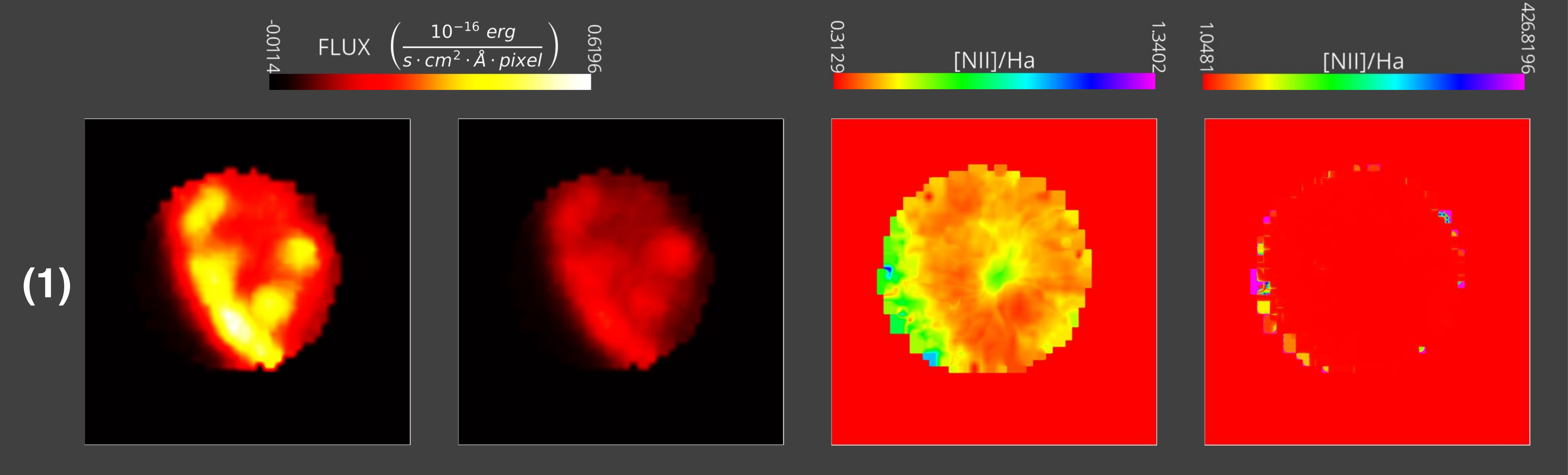}\\
\vspace{-0.05cm}
\includegraphics[width=13cm]{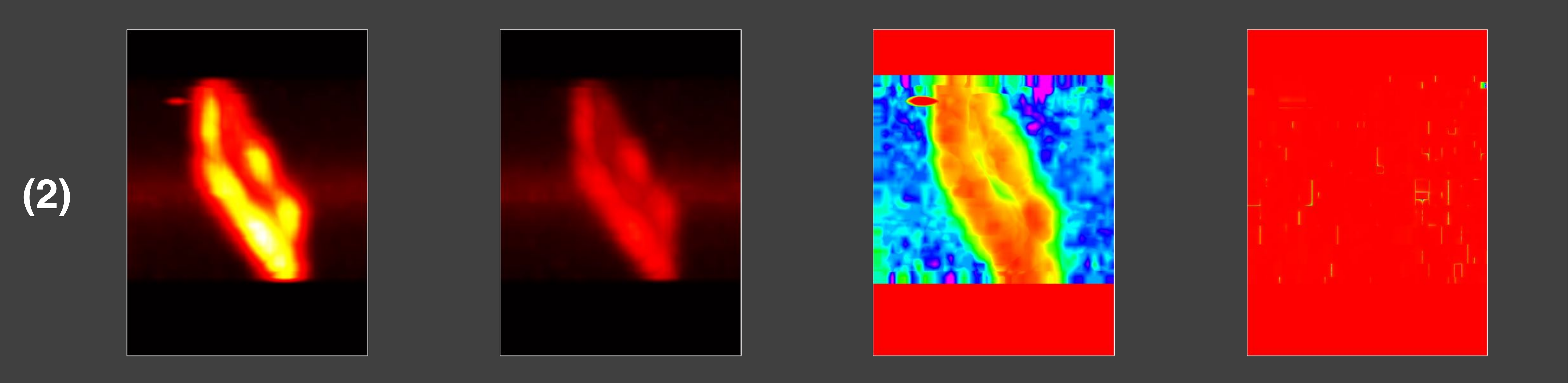}\\
\vspace{-0.05cm}
\includegraphics[width=13cm]{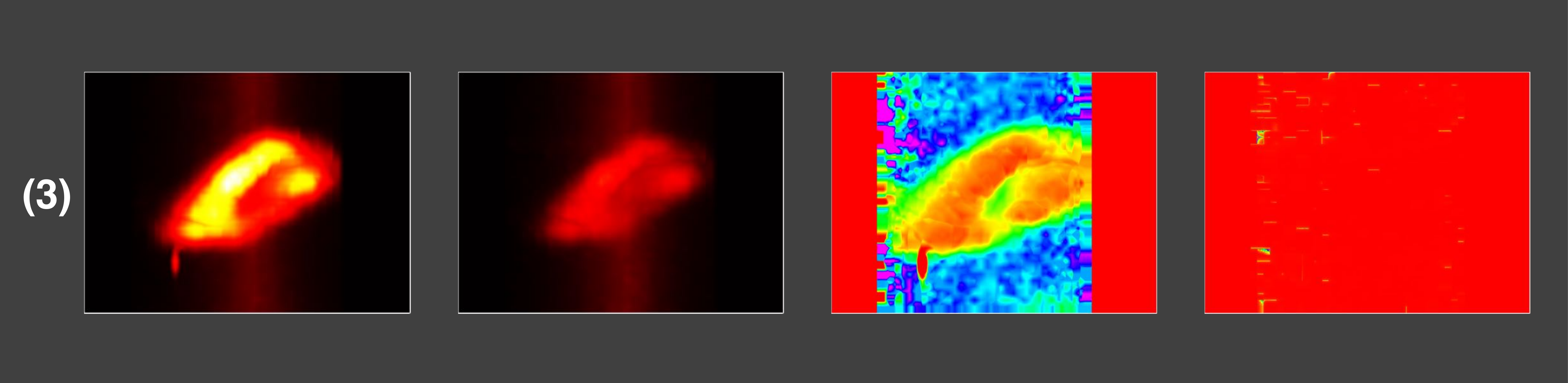}\\
\vspace{-0.05cm}
\includegraphics[width=13cm]{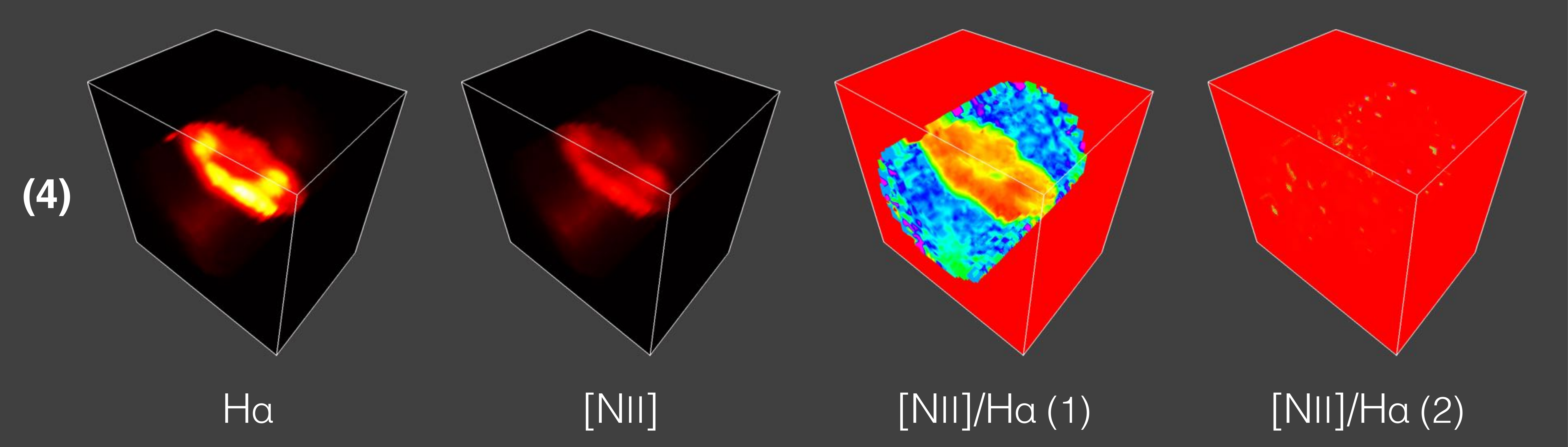}\\
\caption{Computing emission line ratio in fragment shader for the raw data from the H$\alpha$ and the [NII] cubes of GAMA-511867 taken from the SAMI survey. Line ratios are computed on raw data without any filtering or smoothing.}
\label{fig::line-ratio1}
\end{figure*}

\begin{figure*}
\centering
\includegraphics[width=13cm]{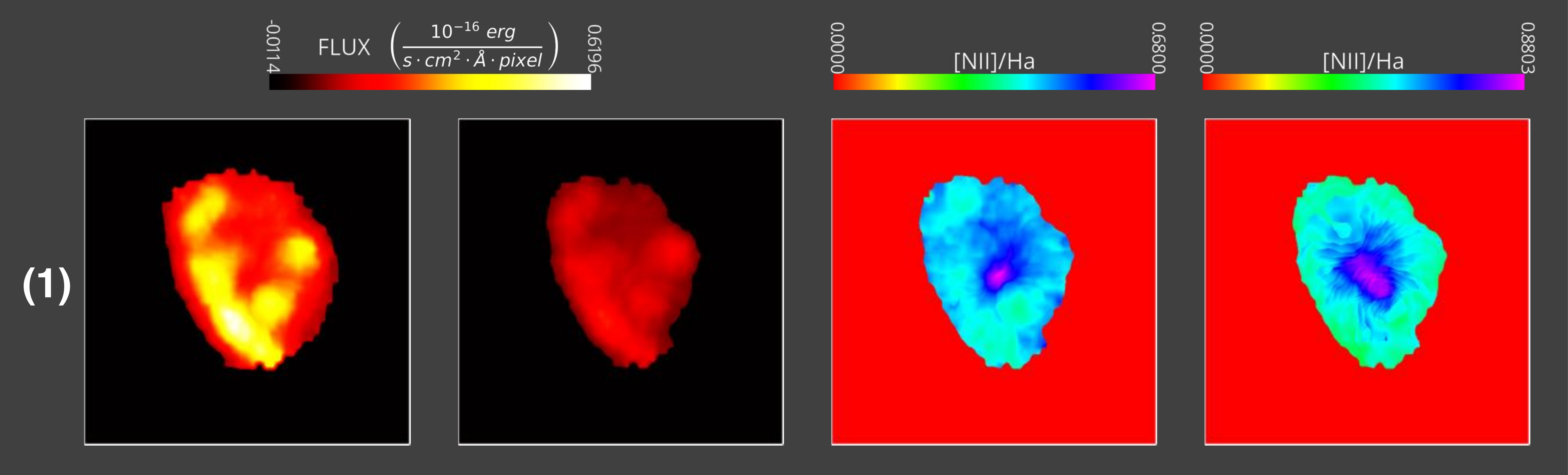}\\
\vspace{-0.05cm}
\includegraphics[width=13cm]{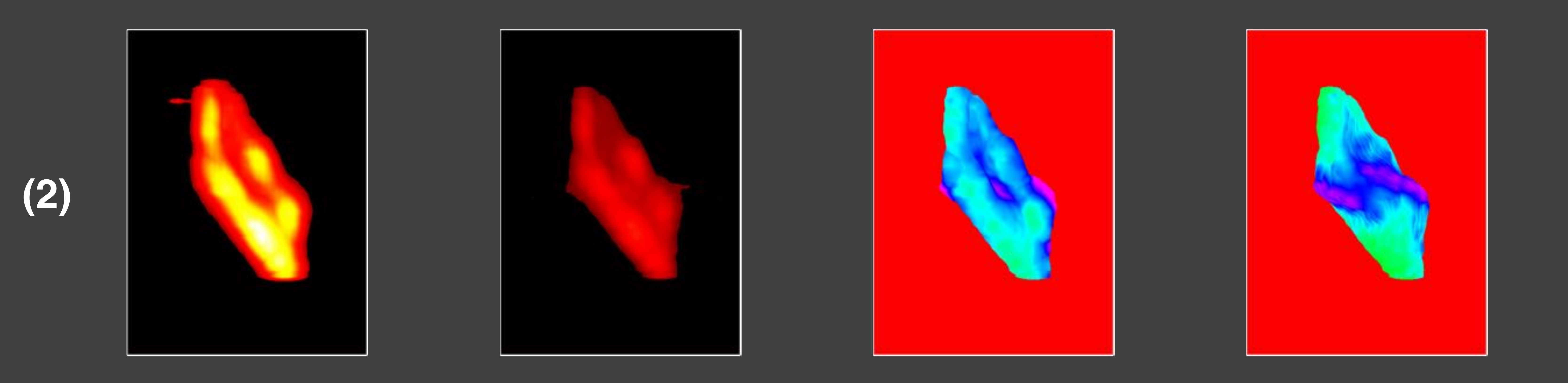}\\
\vspace{-0.05cm}
\includegraphics[width=13cm]{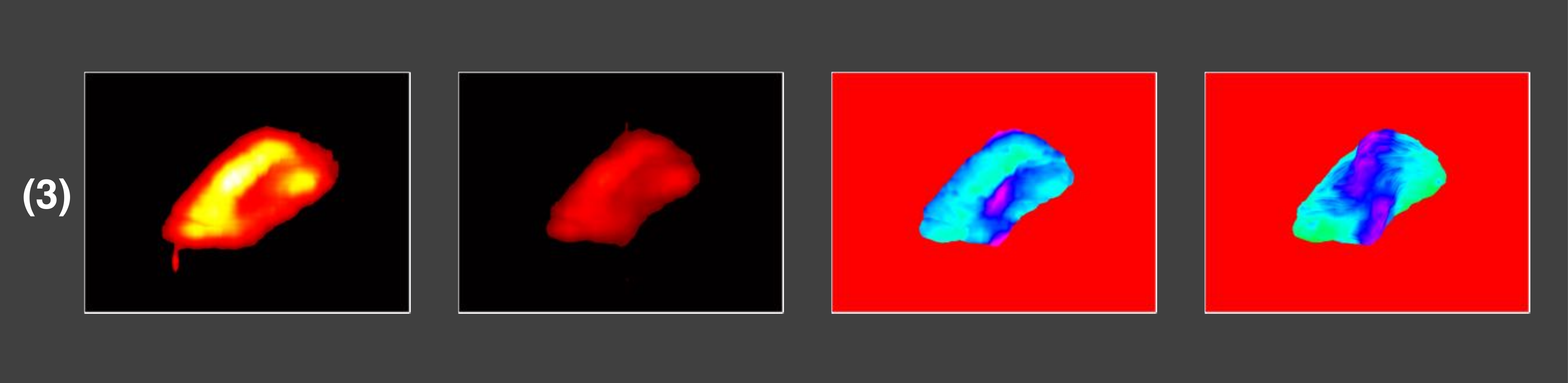}\\
\vspace{-0.05cm}
\includegraphics[width=13cm]{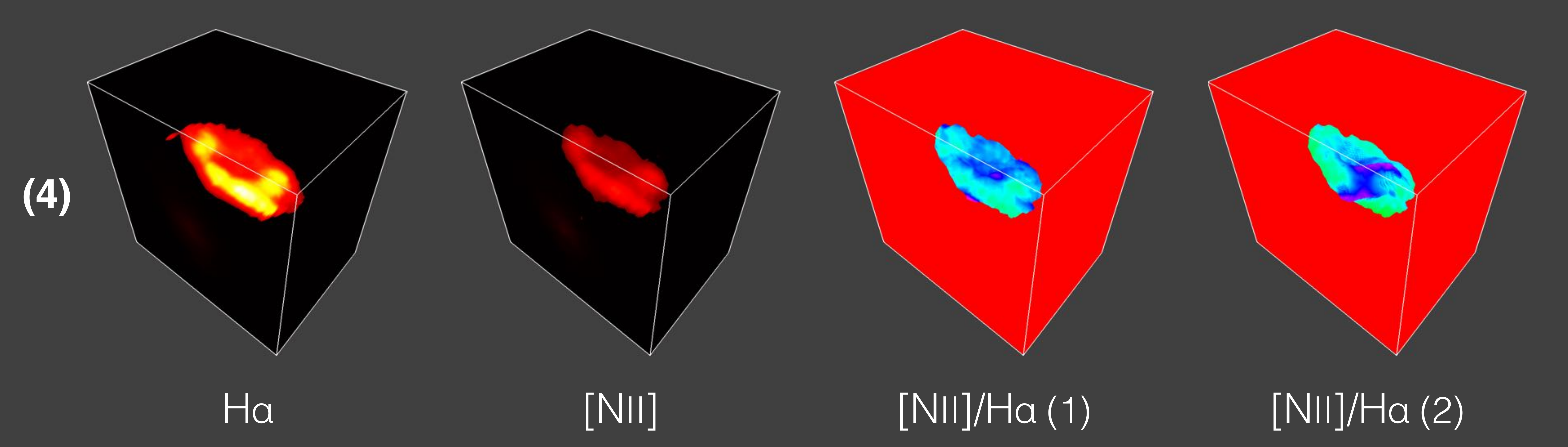}\\
\caption{Computing [NII]/H$\alpha$ in fragment shader for the two sub cubes of GAMA-511867. Panels are arrange as in Figure \ref{fig::line-ratio1}. Filtering using Algorithm $\ref{algo::clipping}$ is applied to discard low flux values from both cubes, using $minT\hspace{-0.1em}hreshold=0.1011$ for the H$\alpha$ cube, and $minT\hspace{-0.1em}hreshold=0.0826$ for the [NII] cube. 
}
\label{fig::line-ratio2}
\end{figure*}

Figures \ref{fig::line-ratio1} and \ref{fig::line-ratio2} show the result of computing the ratio between the [NII] and H$\alpha$ lines of the GAMA-511867 cubes using ray-tracing volume rendering. In both figures, all visualisations use the MIP transfer function. The rows show: 
\begin{enumerate}
\item the spatial view (ra and dec, x and y axes respectively) with parallel projection -- similar to Figure \ref{fig::views}a; 
\item view along spectral (x) and spatial (dec, y) axes; 
\item view along spatial (ra, x) spectral (x) axes; 
\item view showing all three axes with perspective projection. 
\end{enumerate}

From left to right, the different columns show: the H$\alpha$ and the [NII] lines respectively, with flux ranging between 0.4019 to 0.8170 $\times 10^{-16}$ erg s$^{-1}$ cm$^{-2}$ \r{A}$^{-1}$ pixel$^{-1}$; and the emission line ratio computed with techniques from Algorithms \ref{algo::ray-tracing-line-ratio-method1} and \ref{algo::ray-tracing-line-ratio-method2} respectively. Figure \ref{fig::line-ratio1} shows line ratios computed on raw data without any filtering or smoothing. 


As can be seen in Figure \ref{fig::line-ratio1}, faint flux values have an important effect on the visualisation output of the ratio computation. In Figure \ref{fig::line-ratio2}, filtering using Algorithm $\ref{algo::clipping}$ is applied to discard low flux values from both cubes, setting $minT\hspace{-0.1em}hreshold=0.1011$ for the H$\alpha$ cube, and $minT\hspace{-0.1em}hreshold=0.0826$ for the [NII] cube. 
This better constrains the ratio computation to signal from emission lines only, rendering a cleaner visualisation. These values can be selected at runtime.

\subsection{Discussion}
\label{sec::discussion}
Throughout Figure \ref{fig::compare1} to \ref{fig::compare4}, the first row (MIP) highlights an important difference between the MIP$_0$ algorithm and the other two algorithms. In the case of the $MIP_0$, no pixels in the image are transparent. Conversely, with MIP$_1$ and MIP$_{RGB}$, the result of MIP is used to set the transparency level (Algorithm \ref{algo::MIP1}), which makes the rendered image appear fainter. However, this has the ability to provide both information about the voxel intensity and the spectral distribution (e.g. velocity information for first moment-inspired) at the same time. A workaround to tFhis faintness is to only set transparency to values smaller or equal to $minT\hspace{-0.1em}hreshold$. This is shown in Figure \ref{fig::mip-no-alpha}. In the left panel, {\em minT\hspace{-0.1em}hreshold} equals the global minimum, hence no transparency. In the right panel, $minT\hspace{-0.1em}hreshold=1.4\times 10^{-3}$ Jy/beam used with box smoothing ($filterArm=1$). This is an option that can be provided at run-time. 

There is a noticeable difference in the information provided by each transfer function when visualising a blend of spatial and spectral information (Figures \ref{fig::compare2} and \ref{fig::compare4}). The MIP$_0$ and AVIP$_0$ primarily inform about voxel intensity. Even when looking at the snapshots of Figure \ref{fig::compare4}, it can be hard to interpret the velocity structure. MIP$_1$ and AVIP$_1$ clearly inform about velocity, while not being as clear about intensity variation. The RGB shaders show variation in spatial and spectral positions at a glance. For example, when comparing the outcome in row (4) and (5) of Figure \ref{fig::compare2}, each part of the inner and outer spiral arms has a slightly different colour. 

\begin{figure*}
\includegraphics[width=17cm]{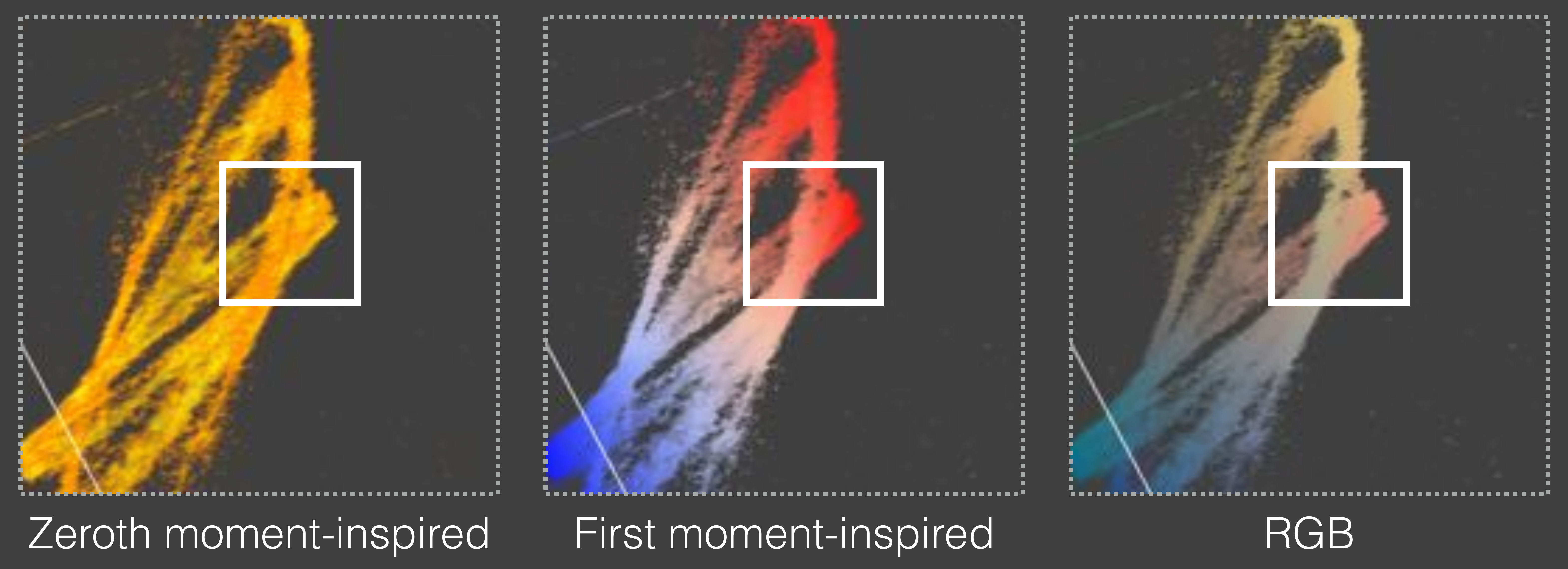}\\
\caption{Close-up on row 5 of Figure \ref{fig::compare2}. The white square shows a region of interest where visual information provided by the colouring technique of each transfer function varies.}
\label{fig::close-up}
\end{figure*}

The use of smoothing can help render ``cleaner'' visualisations for both face-on and side-on views. The direction of the ray tends to have an effect on the result of the ray. For example, when we compare the noise from row (2) in both Figure \ref{fig::compare1} and \ref{fig::compare2}, we can note an accentuation when the cube is viewed at an angle. In this case, rays on different projections through the data sample noise with different correlation properties. For example, noise correlation in the frequency direction is determined by receiver quality, amplifiers, spectral smoothing and binning, etc.; whereas noise correlation in the sky plane is determined, for example, by beam size and properties, and any cleaning previously done to the data. Hence, different lines of sight through the cube will accumulate noise having different correlation properties and so will yield resultant 2D projections with noise characteristics that change with angle. 

In addition, the rendered ``noise'' is likely also related to the step size of the filter kernel and the dimensions of the cubes used in Figures \ref{fig::compare1} to \ref{fig::compare4}. For a face-on sky plane view, the ray will $\geq100$ steps through the data to sample all the voxels in a line of sight, while for a side-on view, the ray will visit $\sim1000$ steps.  Future investigation of adaptive kernel size should help to provide similar noise level at different viewing angles. As can be expected, the 3D kernel of the Gaussian smoothing tends to denoise more than the 1D kernel of the box smoothing (along the ray direction) when the ray traverses the cube at an angle. 

While each method alone may not be required at all times, the fact that shaders are dynamic enables near-instantaneous switching between them at run-time. This provides an important capability to explore different features of the data interactively, and is a clear advantage provided by the use of graphic shaders. 

Consider the following scenario where volume rendering is used along with a common 2D desktop screen. First we visualise the data using the classic zeroth moment-inspired transfer function (Figure \ref{fig::close-up}). While rotating the cube, we reach a viewing angle where information of interest seems to appear (inside the white square in Figure \ref{fig::close-up}). We then want to know where this feature occurs in term of velocity. At run-time, we switch to the first moment-inspired transfer function, which lends extra visual information about the data. Now having an idea of the velocity location of the feature of interest, we want to evaluate its spatial (x,y) location along our line of sight (camera position). Usually, this would be achieved by rotating the object. However, it is now possible to simply switch to the RGB transfer function to acquire this information without having to use rotation. 

The Antennae cube is a good example where the ability to ``boost'' signal using the weighting factor $k$ of AVIP permits a certain intensity level to be highlighted. This can be seen by comparing rows (1) and (2) of both Figures \ref{fig::compare3} and \ref{fig::compare4}. A lower value of $k$ could also have been selected to reduce the saturation, and show internal features in the cube (as in row (5) of Figure \ref{fig::compare2}). Similarly, rows (2) to (4) of Figure \ref{fig::compare4} display how smoothing can reduce the noise and emphasise emission in the data. Figure \ref{fig::compare3} highlights again how the different transfer functions can provide information about the overall distribution (moment 0) and velocity field (moment 1). Clumps of emission are clearly visible throughout the visualisation. Both smoothing methods provide similar visual outcome --- where the Gaussian smoothing does not discard values as drastically as the large box smoothing ($filterArm=5$). 

Finally, the visualisation of the [NII]/H$\alpha$ line ratio highlights an important feature enabled by graphics shaders. In addition to the development of custom transfer functions, the shader can be further used to compute physical information about datasets. Here again, being able to compute these quantities at run-time provides great potential compared to traditional pre-processing methods. Having the whole dataset available enables interactive searches for interesting features by mixing different algorithms together (e.g. filtering and ratio). This is important in the case of the voxel by voxel computation, for which purely computing the ratio on raw data produces highly divergent bounds --- rendering very few visual features. Intensity domain scaling (Algorithm \ref{algo::scale}) could also provide a dynamic way to look for interesting features within these divergent bounds.

While line ratio visualisation is generally done with a face-on view (ra and dec), in Figures \ref{fig::line-ratio1} and \ref{fig::line-ratio2} we also show different viewing angles in both Figures. Being able to rotate the cube promotes improved understanding of which component of the 3D structure produces the features in the face-on view.  We repeat that these visualisation are used as a proof of concept, and that more effort is required to provide high precision line ratio maps --- including a fitting procedure to reduce the inherent noisiness of the data.

\section{Performance}
\label{sec::performance}

Apart from the visual outcome of the transfer functions, an important aspect of the shaders relates to their algorithmic complexity. Shader algorithms are computed at run-time on the GPU: no pre-computing is required. Each ray is processed in parallel in GPU memory. It was mentioned in Section \ref{sec::transfer_functions} that both MIP and AVIP have a complexity of $O(N)$ per line of sight, where all voxels along the line of sight (or more precisely the sampled voxels) are visited once to evaluate the maximal value or the weighted average respectively. Similarly, computing smoothing as part of the shaders adds to the initial $O(N)$ by increasing the number of voxels visited by a ray. To evaluate the effect of smoothing on frame rate, we perform two timing benchmarks. Details about the data and hardware configuration used in our benchmark can be found in Section Appendix \ref{app:data-hardware}.

For our first benchmark, we evaluate the frame rate (in frames per second) computed on the GPU during visualisation for different smoothing parameters (e.g. 9-tap Gaussian smoothing, 13-tap Gaussian smoothing, ...), using both MIP and AVIP. The visualisation is rendered into a canvas of $1000 \times 1000$ pixels. We record the frame rate during a full rotation of the cube. We repeat this process three times for each kernel, and report the median frame rate.

\begin{figure*}
\centering
\includegraphics[width=17cm]{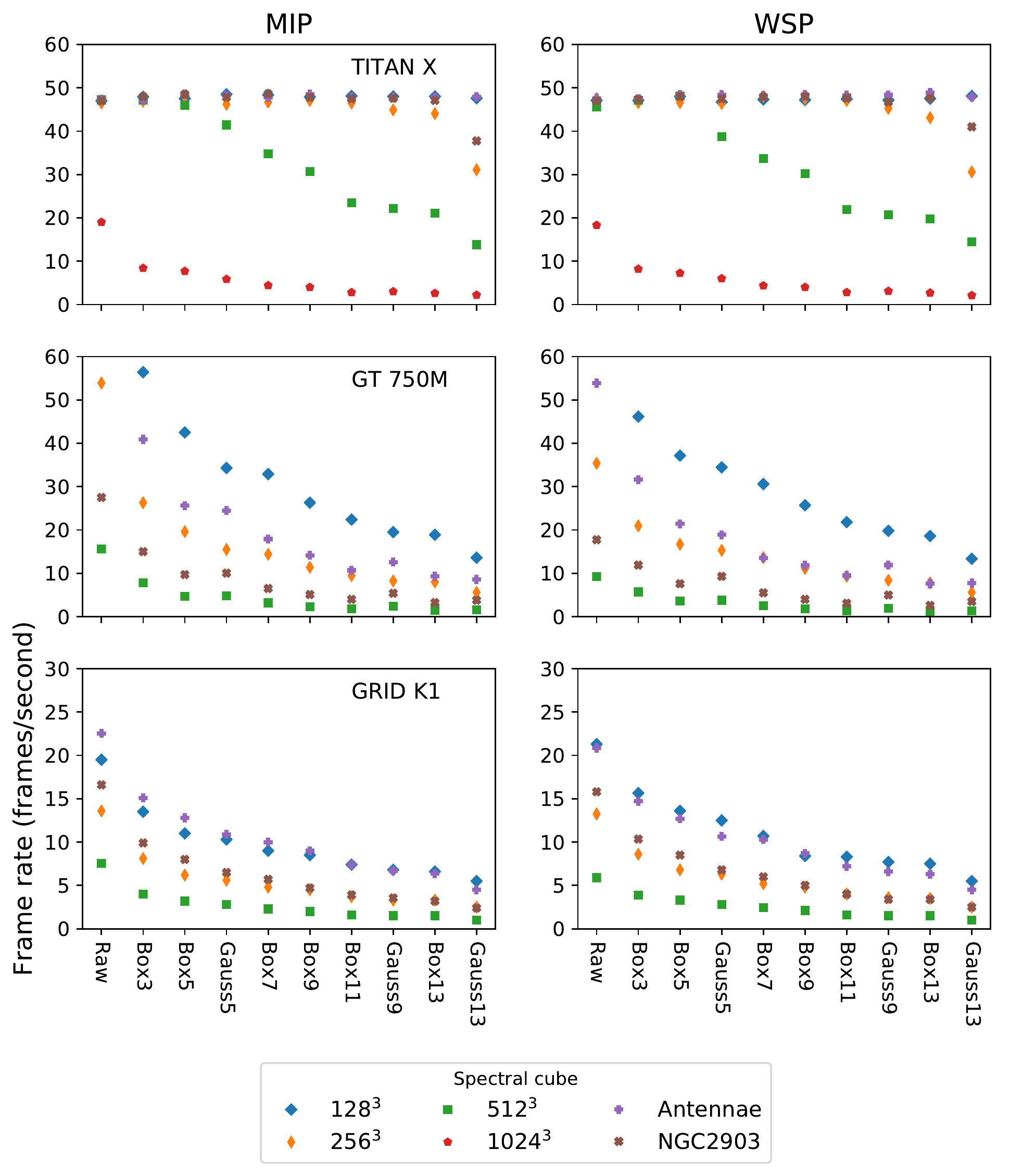}
\caption{Median frame rate (frames per second) for three rotation of a spectral cube. Each row presents the FPS obtained on different GPUs, indicated inside the left panel (TITAN X, GT 750M, GRID K1). The left and right columns show the results for MIP and AVIP respectively. The smoothing parameter are ordered as a function of the number of texture fetch required by the smoothing shader. Raw indicates no smoothing. Each spectral cube evaluated is indicated in the legend on the right of the figure.}

\label{fig::fps}
\end{figure*}

We report the benchmark's results in Figures \ref{fig::fps}. We sort the parameters as a function of the number of texture fetches for each step of the ray. In our implementation, for an $N$-box filter, the number of texture fetches [e.g. $cube(loc+i)$ in the algorithms] is $N$; for the $N$-tap 3D Gaussian kernel, the number of texture fetches is $3\times\frac{N-1}{2}+1$. Note that we report results for the $1024^{3}$ cube only for the TITAN X GPU as median frame rates for the other GPUs were $\sim2$ frames per second without smoothing. Without smoothing (Raw), each step of the ray fetches a single voxel.

From Figure \ref{fig::fps}, we can first note that all GPUs are affected by the data size. For all GPUs, the largest cube is the one with the slowest frame rate. As the size of the data increases, each ray needs to visit an increasing number of voxels to compute the pixel value. Additionally, once loaded in memory, less memory is available for computation -- slowing down the rate at which frames are generated. We can also note a reduction in frame rate as a function of the number of texture fetches. As expected, AVIP and MIP show similar frame rates, with AVIP being slightly slower. The ``High-end'' GPU (TITAN X) presents more sustained frame rates (with the exception of the largest cubes) than ``low-end'' GPUs like the GT 750M. This can be attributed in part to the larger memory available, the number of processing cores, and faster clocks. In some cases however, GT 750M was faster than TITAN X, which peaked at 94 frames per second for the $128^3$ cube with MIP-Raw (cropped out of the figure). We suspect this may be due to caching capabilities of macOS. 

In Figure \ref{fig::fps}, the panels related to the GRID K1 have a different vertical axis range than the other two. The drop in frame rate can primarily be attributed to the use of VirtualGL, which converts frames using the JPEG still image compression. VirtualGL is known to offer limited speed for very high resolutions canvases \citep{Lietsch2008, Lohnhardt20105693145}. It is possible that the slightly slower clock of the GRID K1 also plays a part in the lower frame rate. Nevertheless, this benchmark shows that is is possible to visualise spectral cube remotely on the Cloud. This pathway -- which is likely to become better with time (better hardware and internet connections) -- could provide a good option for users, removing the need to install software or to buy and maintain expensive hardware locally.

\begin{figure*}
\centering
\includegraphics[width=17cm]{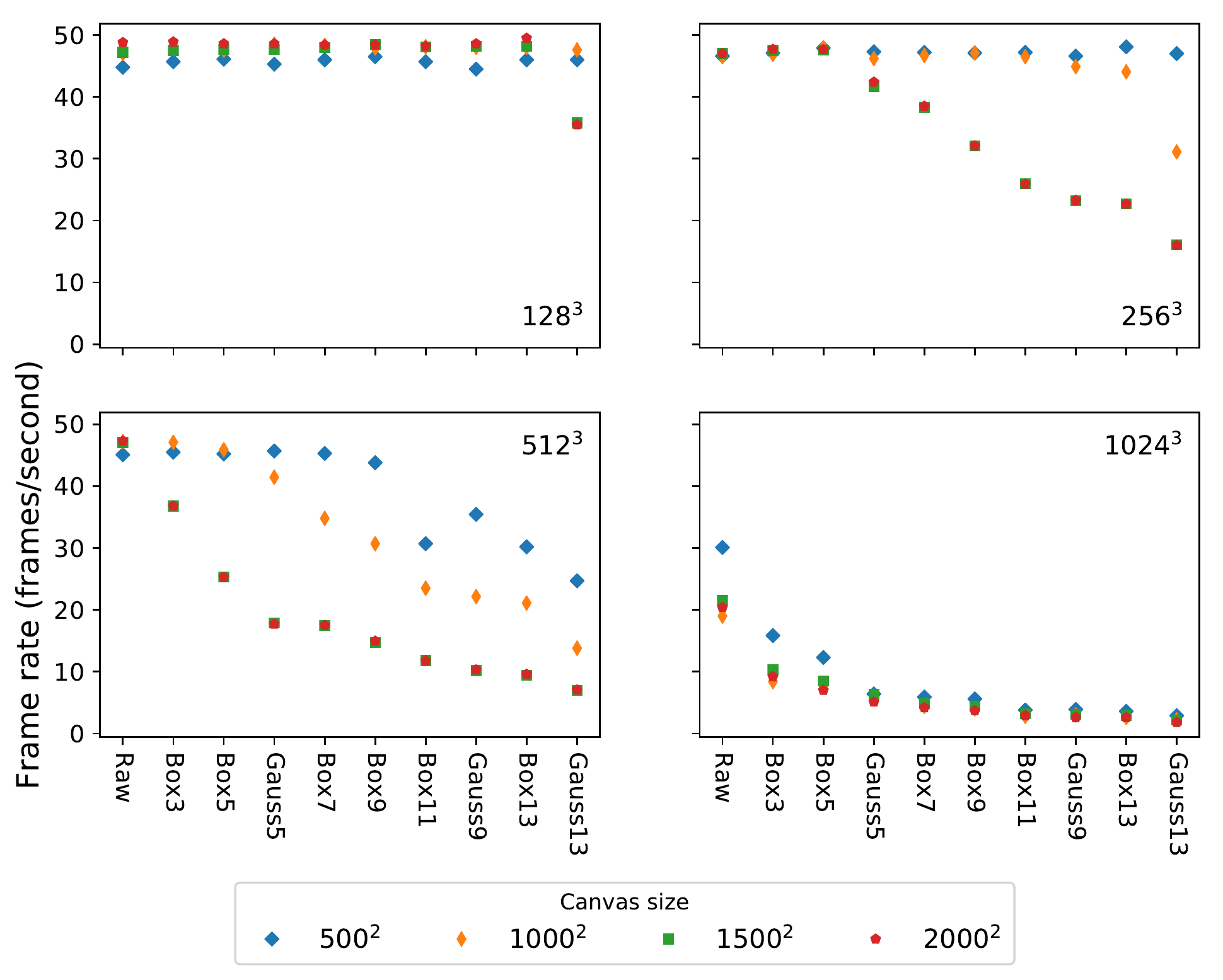}
\caption{Comparing MIP's frame rate (frames per second) at different canvas sizes using TITAN X for the $128^{3}$, $256^{3}$, $512^{3}$, and $1024^{3}$ cubes.}

\label{fig::canvassize}
\end{figure*}

For our second benchmark, we evaluate the effect of canvas size on frame rate using our best GPU (TITAN X). In this case, we record the frame rate with canvas size varying from $500^{2}$, $1000^{2}$, $1500^{2}$, to $2000^{2}$ pixels using the $128^{3}$, $256^{3}$, $512^{3}$, and $1024^{3}$ cubes. We present the results in Figure \ref{fig::canvassize}. For our smallest dataset, canvas size has a very limited effect, showing only small variations in frame rate. For larger datasets, results show that canvas size has an effect on frame rate. In addition to the effect of data size, the slowdown can be attributed to the fact that, as the canvas size expands, a larger number of pixels need to be drawn at each time step. While the code has not been completely optimized, this shows that we are reaching physical limits of a single GPU. If real-time interactivity and high frame rates are needed for larger spectral cubes, our solution could be integrated with distributed rendering techniques like those introduced by \citet{Hassan2011100} and \citet{Hassan2013MNRAS.429.2442H}.

With a canvas with resolution of about full HD ($1920 \times 1080$), the TITAN X GPU can render all smoothing kernels considered in this paper at 15 frames per second or more for data sizes up to $256^{3}$, and up to a box kernel size of 9 ($filterArm=4$) and a 5-tap Gaussian kernel for data size of $512^{3}$.

From these results, we can conclude that smoothing data can be done at run time with limited effect on interactivity. The interactivity can play a great role during data exploration, as it can be difficult to select a set of parameters {\em ad-hoc}. Having the possibility to explore and manipulate data in realtime represents a powerful tool for the user -- a step forward for the development of next-generation visualisation and analysis software. 

\section{Conclusion and future work}
\label{sec::conclusion}
As current and upcoming instruments and facilities allow larger and more numerous spectral cubes to be gathered, there is a need for novel and more efficient visualisation techniques to be explored. Ray-tracing volume rendering allows astronomers to inspect spectral cubes as a whole --- a step beyond visualisations like channel and moment maps. However, transfer functions classically used with ray-tracing volume rendering only provide information about the overall intensity distribution --- similar to the zeroth moment map. 

For the first time, this article presented transfer functions going beyond overall intensity by adding visual cues to spatial and velocity information in 3D space (see for example Figures \ref{fig::mip-no-alpha} and \ref{fig::close-up}). We described a transfer function that produces the 3D equivalent of the first moment maps, providing a quick visual cue about the velocity field. Furthermore, we presented a generalization of this first moment transfer function, based on the RGB cube, informing about all three dimensions at once (spatial and spectral dimensions). We compared the outcome of all transfer functions using HI and CO spectral cube data. We also presented methods to compute an emission line ratio with ray-tracing volume rendering. Using two sub-cubes of H$\alpha$ and [NII] taken from an IFU observation, these methods shows that graphics shaders can be utilised to further compute physical information.

We also showed that the GPU and graphics shaders can be utilised to provide fast computation of these transfer functions. In particular, we used the fragment shader --- part of the graphics pipeline dedicated to colouring pixels in the final image --- to compute our transfer functions efficiently. In addition, we showed that common pre-processing algorithms such as filtering and smoothing (e.g. box and Gaussian smoothing) can be computed on-the-fly in the fragment shader. This approach opens new ways to interactively explore spectral cubes in order to find parameters of interest to be used in further quantitative investigation of the data, such as smoothing kernel size for source finding. Future work should investigate automation of parameters selection (e.g. $k$) and their relation to physical quantities. 

The transfer functions presented in this article highlight an important aspect of the computation of moment maps: as Equations \ref{eq::mom0} and  \ref{eq::mom1} have to be computed for every line of sight (say $M$ times), the overall complexity of the algorithm is $O(MN)$, where $M$ is the number of pixel in the image, and $N$ is the number of velocity or spectral channels. It qualifies as what is generally called an ``embarrassingly parallel problem''. By utilizing the thousands of parallel cores offered by modern GPUs, the transfer functions are fast to the point that for the dataset evaluated in this work, the computation of the zeroth and first moment maps can be reduced to an algorithmic complexity $O(N)$, a considerable improvement.

At Full HD resolution ($1920 \times 1080$), high-end GPUs ($\sim USD\$1000$) can render all smoothing kernels considered in this paper  -- 1D Box  and 3D Gaussian smoothing -- at 15 frames per second or more for data sizes up to $256^{3}$, and slightly smaller kernels for data of size $512^{3}$. 

Our investigation suggests that custom transfer functions and shaders can play an important role in the development of future visualisation and analysis astronomical software. 

\section*{Acknowledgements}

This research was supported by use of the NeCTAR Research Cloud and by the Melbourne Node at the University of Melbourne. The authors thank Bernard Meade for providing the NeCTAR node image. The NeCTAR Research Cloud is a collaborative Australian research platform supported by the National Collaborative Research Infrastructure Strategy. This work made use of THINGS, `The HI Nearby Galaxy Survey' \citep{Walter2008AJ}. This paper makes use of the following ALMA data: ADS/JAO.ALMA\#2011.0.00003.SV. ALMA is a partnership of ESO (representing its member states), NSF (USA) and NINS (Japan), together with NRC (Canada), NSC and ASIAA (Taiwan), and KASI (Republic of Korea), in cooperation with the Republic of Chile. The Joint ALMA Observatory is operated by ESO, AUI/NRAO and NAOJ. The SAMI Galaxy Survey is based on observations made at the Anglo-Australian Telescope. The Sydney-AAO Multi-object Integral field spectrograph (SAMI) was developed jointly by the University of Sydney and the Australian Astronomical Observatory. The SAMI input catalogue is based on data taken from the Sloan Digital Sky Survey, the GAMA Survey and the VST ATLAS Survey. The SAMI Galaxy Survey is funded by the Australian Research Council Centre of Excellence for All-sky Astrophysics (CAASTRO), through project number CE110001020, and other participating institutions. The SAMI Galaxy Survey website is \url{http://sami-survey.org/}. We thank the anonymous reviewer for prompting us to develop the line ratio shader. DV thanks the {\em Fonds de recherche du Qu\'{e}bec -- Nature et technologies} (FRQNT) and Swinburne Research for postgraduate scholarships.




\bibliographystyle{mnras}
\bibliography{bib-coloring_velocity} 




\appendix
\section{Projections for ray-tracing volume rendering}
\label{app::projections}
\begin{figure}
\includegraphics[width=8.3cm]{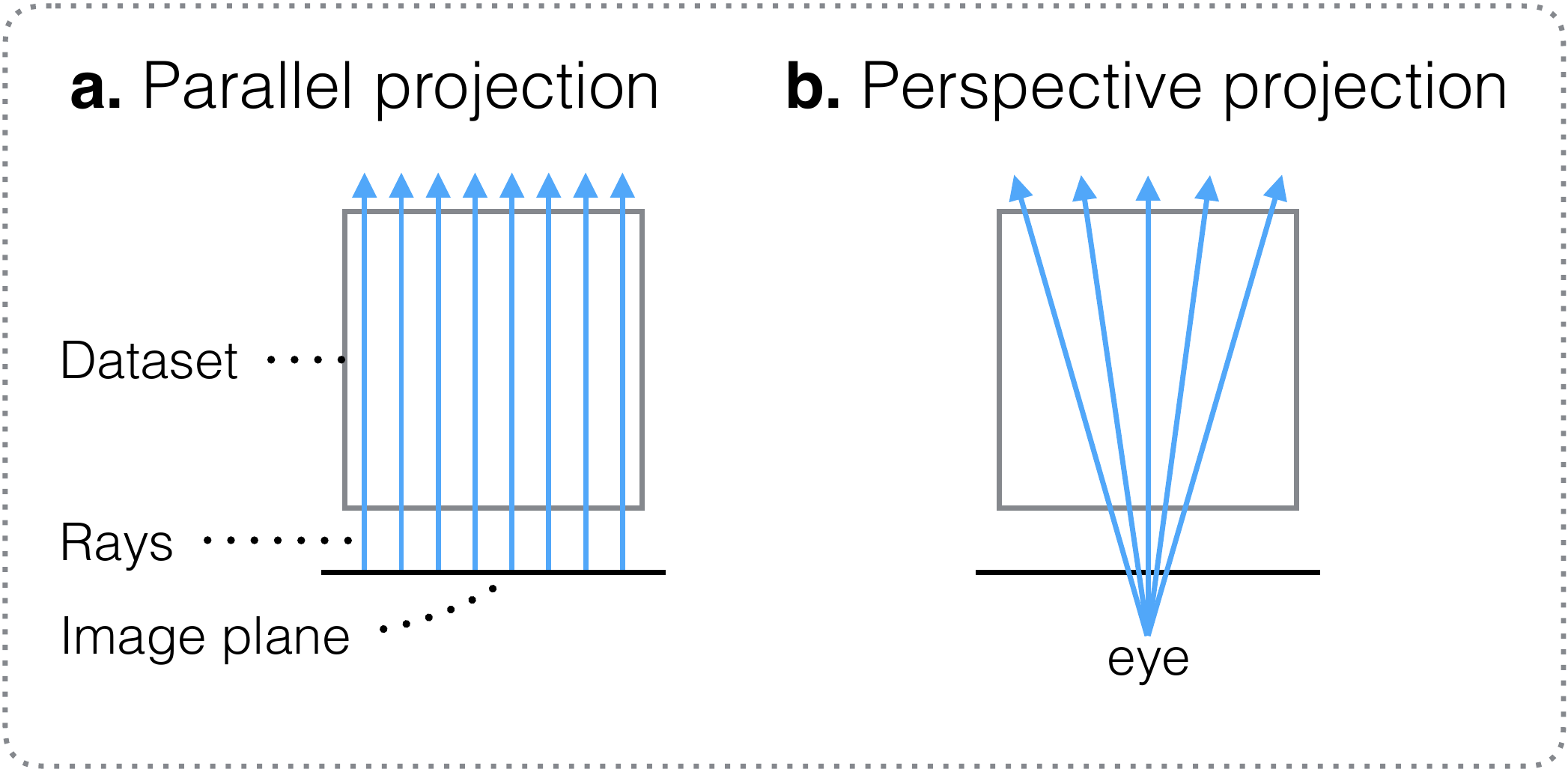}
\caption{Common projections used with the ray-tracing technique: a. Parallel projection, where each ray is parallel to the next; and b. Perspective projection, where all rays are shot from a focal point (e.g. eye of the viewer).}
\label{fig::projections}
\end{figure}

\begin{figure}
\includegraphics[width=8.3cm]{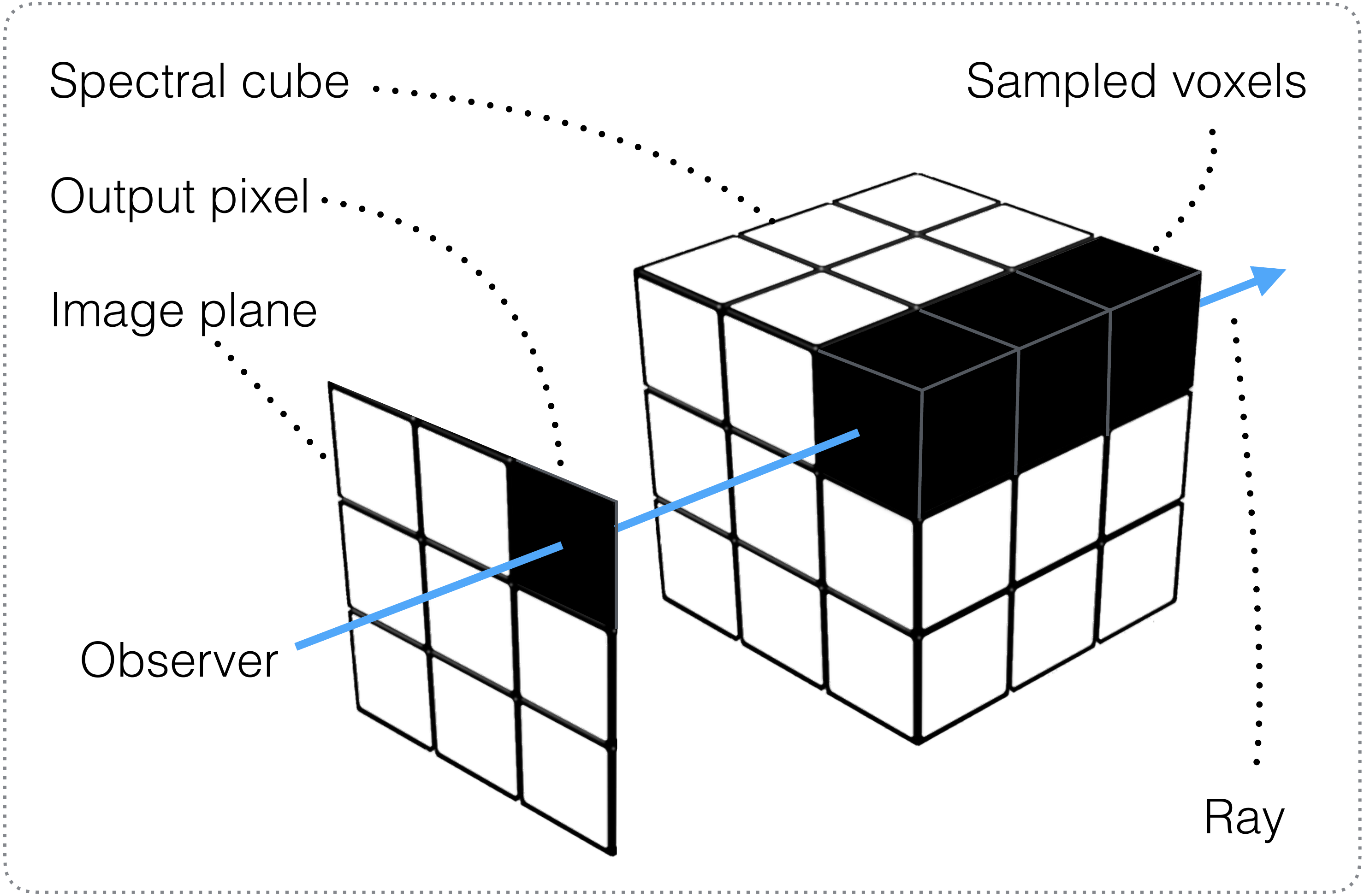}
\caption{Schematic of the ray-tracing technique, assuming a parallel projection. {\em Ray-tracing} calculates pixel values in the final image based on voxel values encountered while ``shooting'' rays through the spectral cube. The {\em transfer function} combines the voxels encountered by the ray into the final image's pixel colour and transparency values. }
\label{fig::raytracing}
\end{figure}

In ray-tracing volume rendering, rays are traced using either a parallel projection (also known as orthographic projection) or a perspective projection. For a parallel projection, each ray has a trajectory that is parallel to other rays. For a perspective projection, all ray trajectories diverge from a focal point that corresponds to the observer's eye. Figure \ref{fig::projections} depicts rays as traced by both projection type. Figure \ref{fig::raytracing} shows a schematic of the ray-tracing technique using parallel projection. 

 \section{Data and hardware description for performance experiment}
 \label{app:data-hardware}
 \begin{table*}
\centering
\caption{GPUs and other hardware description used for benchmark.}
\label{table::gpus}
\begin{tabular}{@{}lll@{}lll@{}}
\hline
\hline
\textbf{GPU}			& \textbf{Environment}  \\ 
\hline
{\bf NVIDIA GeForce GTX TITAN X}	& Desktop computer	(Windows 7) 	 \\
12 GB of GDDR5	 & Intel Xeon CPU ES-2609 v2 at 2.5 GHz\\
GPU clock :  1.00 GHz			& 16 GB RAM at 2.5 GHz	  	 \\
Boost Clock: 1.09 GHz			& Sony VPL-VW100ES (4K projector) \\
							& Display resolution : $3840 \times 2160$ \\

\hline
{\bf NVIDIA GeForce GT 750M}		& Laptop computer (macOS Sierra 10.12.3) 	\\
2 GB of DDR3/GDDR5			& Intel Core i7 at 2.5 GHz CPU		\\
GPU clock:  941 MHz			& 16 GB RAM at 1.6 GHz	\\
Boost clock: 967 MHz			&  Dell E2313H (external monitor) \\
							&  Display resolution : $1920 \times 1080$\\

\hline
{\bf NVIDIA GRID K1}			& Remote Desktop [Ubuntu (MATE Desktop Environment 1.8.2)] \\
4 $\times$ 4GB of DDR3			& Intel QuadCore Haswell CPU 	 \\
GPU clock: 850 MHz				& 16 GB of RAM at 2 GHz		  \\
							& HP compaq LA2405wg monitor \\
							& Remote display resolution : $1920 \times 1074$ \\
\hline
\hline
\end{tabular}
\end{table*}
 
We evaluate the frame rate for a number of spectral cubes. First, we generated cubes ranging from $128^3$ to $1024^3$ voxels in size, for which each voxel value is a random 32-bit floating point value ranging from [0, 1]. For these cubes, when using AVIP, we set $k=0.22$ and $minT\hspace{-0.1em}hreshold=0.9$. Secondly, we evaluate cubes that are not equal in all three dimensions: the NGC2903 and Antennae cubes for which dimensions are already listed in Section \ref{sec::testdata}. For these two cubes, when using AVIP, we set $k=0.22$ and $minT\hspace{-0.1em}hreshold=7\times10^{-3}$ Jy/beam. The value of $minT\hspace{-0.1em}hreshold$ is chosen to minimize the effect of the exit strategy (when $\alpha>0.99$) of AVIP to provide a fair comparison with MIP.

We performed the same benchmark on three different GPUs: NVIDIA GeForce GTX TITAN X, NVIDIA GeForce GT 750M, and NVIDIA GRID K1. Table \ref{table::gpus} presents the specification of each GPU and their respective host environments. The TITAN X and GT 750M GPUs are hosted on local computers, while the GRID K1 is hosted on The National eResearch Collaboration Tools and Resources project (Nectar) Research Cloud\footnote{The Nectar Research Cloud is an online infrastructure that includes software and services allowing Australia's research community to store, access, and run data remotely. Details can be found at \url{https://nectar.org.au}.}, and accessed using a remote desktop configuration. We access the remote desktop via a desktop computer (CentOS 6.7) equipped with 16 GB of RAM and an NVIDIA Geforce GTX 470. In the remote desktop setting, the software is run using VirtualGL\footnote{\url{http://www.virtualgl.org}} ({\tt vglrun}) over TurboVNC\footnote{\url{http://www.turbovnc.org}}.




\bsp	
\label{lastpage}
\end{document}